\documentclass[prd,aps,showpacs,notitlepage,superscriptaddress,nofootinbib]{revtex4-1}

\usepackage{epsfig}
\usepackage{color}
\usepackage{url}
\usepackage{amsmath}
\usepackage{multirow}

\def\bea#1\eea{\begin{align}#1\end{align}}

\makeatletter
\def\slash#1{{\mathpalette\c@ncel{#1}}} % TeXbook, bottom of p360
\makeatother

\newcommand\beq{\begin{eqnarray}}
\newcommand\eeq{\end{eqnarray}}

\newcommand\la{\langle}
\newcommand\ra{\rangle}

\begin{document}
\title{New approach to the Sivers effect in the collinear twist-3 formalism}

\date{\today}
                       
\author{Hongxi Xing}
\email{hxing@m.scnu.edu.cn}
\affiliation{Institute of Quantum Matter and School of Physics and Telecommunication Engineering,
South China Normal University, Guangzhou 510006, China} 

\author{Shinsuke Yoshida}
\email{shinyoshida85@gmail.com}
\affiliation{Institute of Quantum Matter and School of Physics and Telecommunication Engineering,
South China Normal University, Guangzhou 510006, China}

\begin{abstract}

The single-transverse spin asymmetry(SSA) for hadron production in the transversely polarized proton scattering receives major contribution from Sivers effect, which can be systematically described within the collinear twist-3 factorization framework in various processes. Conventional method in the evaluation of the Sivers effect known as pole calculation is technically quite different from non-pole calculation which is another method used in evaluating the final state twist-3 effect. In this paper, we extend the non-pole technique to the Sivers effect, and show the consistency with the conventional method through an explicit calculation of $\mathcal{O}(\alpha_s)$ correction in semi-inclusive deep inelastic scattering. As a result, we clarify that the conventional pole calculation is implicitly using the equation of motion and the Lorentz invariant relations whose importance became widely known in the non-pole calculation. We  also clarify some technical advantages in using the new non-pole method.

\end{abstract}

\maketitle

\section{Introduction}

The origin of the single transverse-spin asymmetries(SSAs) in high-energy hadron scatterings has been a long-standing mystery over 40 years since the strikingly large asymmetries were observed in mid-1970s~\cite{Klem:1976ui,Bunce:1976yb}. RHIC experiment has provided many data of the SSAs for various hadron productions in the last decade~[3-7] and motivated a lot of theoretical work on the development of the perturbative QCD framework. Much theoretical effort has been devoted to develop a reliable QCD-based theory in order to deal with the given experimental data. The twist-3 framework in the collinear factorization approach was established as a rigorous framework which can provide a systematic description of the large SSAs. 

It is commonly known that there are two major effects which lead to the large SSAs observed in the experiment, i.e., initial state Sivers effect and final state Collins effect. The Sivers effect is essentially twist-3 contribution generated from a transversely polarized hadron in the initial state. Started from the pioneering work by Efremov and Teryaev~\cite{Efremov:1981sh}, more systematic techniques were developed in a series of studies around '00~[9-12]. A solid theoretical foundation for the calculation of the Sivers effect was finally laid in Ref.~\cite{Eguchi:2006mc}. We will show the calculation technique in detail in next section and here we just give a brief introduction. The Sivers effect of the transversely polarized proton can be expressed by the dynamical twist-3 function 
defined by a Fourier transformed proton matrix element
$T_{q,F}\simeq {\cal F. T.}\la pS_{\perp}|\bar{\psi}gF^{+-}\psi|pS_{\perp}\ra$ and the cross section in deep inelastic scattering (DIS) can be derived as
\beq
d\sigma=iT_{q,F}\otimes D \otimes d\hat{\sigma},
\label{twist-3X}
\eeq
where $D$ represents the usual twist-2 fragmentation function and $d\hat{\sigma}$ is a hard partonic cross section. Because all the nonperturbative functions are real in this equation, the partonic cross section has to give an imaginary contribution in order to cancel $i$ in the coefficient. This imaginary contribution can be given by the pole part of a propagator in the partonic scattering. In the quantum field theory, the propagator is defined by the time-ordered product of two fields and it has $i\epsilon$-term in the denominator. The imaginary contribution can emerge from a residue of contour integration. This is a basic mechanism of the pole calculation for the Sivers type contribution. Next we turn to the twist-3 fragmentation effect of spin-0 hadron which is known as Collins effect. The cross section formula for the twist-3 fragmentation contribution was completed in $pp$ collision~\cite{Metz:2012ct} and DIS ~\cite{Kanazawa:2013uia} in a formal way. The dynamical twist-3 fragmentation function can be defined as $\hat{D}_{q,F}\simeq {\cal F. T.}\la 0|gF^{+-}\psi|hX\ra\la hX|\bar{\psi}|0\ra$ and the cross section in DIS is expressed by the same form as Eq. (\ref{twist-3X}) just replacing $T_{q,F}$ with $\hat{D}_{q,F}$ and $D$ with the usual twist-2 parton distribution function respectively. The main difference is that the fragmentation function $\hat{D}_{q,F}$ is complex and therefore it gives the imaginary contribution. The pole contribution from the hard cross section is no longer needed. This is a mechanism of the non-pole contribution from the Collins effect. The origin of generating the imaginary phase results in the main technical difference between the calculations for the Sivers and the Collins effects.  
The cross section for the pole contribution only depends on the dynamical function, while the result for the non-pole contribution is expressed in terms of three types of nonperturbative functions, the dynamical, intrinsic and kinematical functions. In general, the hard cross sections in the non-pole calculation are not gauge- and Lorentz-invariant, because they are not physical observables, only their sum leads to physical result as measured by experiment. This problem is solved by using two types of the relations among the nonperturbative functions which are called equation of motion relation and Lorentz invariant relation~\cite{Kanazawa:2015ajw}. 

As discussed above, the calculations for the pole contribution and the non-pole contribution are technically different from each other. Although the calculation techniques for those contributions are important basics of the higher twist calculation, not so many theorists are familiar with both of them because of the technical differences. In this paper, we revisit the result of the pole calculation from the viewpoint of the non-pole calculation in order to understand two calculations in a unified way. We show that the non-pole calculation has several technical advantages, thus should be extended to more complicated calculations like next-to-leading order calculation and twist-4 calculation.

%various channels and higher twist calculations.

The remainder of the paper is organized as follows: in Sec.~II we introduce the notation and review the conventional pole calculation in detail. In Sec.~III we show the non-pole calculation method for the twist-3 contribution in order to reexamine
the pole contributions. Finally, in Sec.~IV we summarize the achievements in this paper and make some comments on possible applications of the new non-pole method.

%-----------------------   Section 2 ----------------------

\section{Conventional pole calculation at twist-3}

The conventional collinear expansion framework at twist-3 has been developed in Refs.~[8-17]. We review here the pole calculation for semi-inclusive deep inelastic scattering (SIDIS) in order to clarify the difference from the new method of non-pole calculation that we will propose in next section. 
SIDIS is a suitable process to check the consistency between the two methods because the twist-3 cross section has been already completed in Ref.~\cite{Eguchi:2006mc} based on the conventional method. 
In addition, SIDIS is a relatively easier process than $pp$-collision because of some technical issue.  %We will discuss this issue after introducing Ward-Takahashi identities in Sec. III.

We consider the process of polarized SIDIS 
\beq
e(l)+p^{\uparrow}(p,S_{\perp})\to e(l')+h(P_h)+X,
\eeq
where the initial proton is transversely polarized. $l$ and $l'$ are, respectively, the momenta of the incoming and outgoing electrons. $p$ and $S_{\perp}$ are the momentum and the transverse spin of the beam proton, $P_h$ is the momentum of the final state hadron.
In this paper, we focus on one-photon exchange process with the photon invariant mass $q^2=(l-l')^2=-Q^2$, the extension to charged current interaction is straightforward. The polarized cross section for SIDIS is given by
\beq
{d^4\Delta\sigma\over dx_Bdydz_hdP_{h\perp}}={\alpha^2_{em}\over 32\pi^2z_hx_B^2S^2_{ep}Q^2}L^{\mu\nu}W_{\mu\nu},
\eeq
where the standard Lorentz invariant variables in SIDIS are defined as
\beq
S_{ep}=(p+l)^2,\hspace{5mm}x_B={Q^2\over 2p\cdot q},\hspace{5mm}
z_h={p\cdot P_h\over p\cdot q},\hspace{5mm}y=\frac{p\cdot q}{p\cdot l}.
\eeq
The leptonic tensor is defined as follows:
\beq
L^{\mu\nu} = 2\left(l^{\mu}l'^{\nu}+l^{\nu}l'^{\mu}-{Q^2\over 2}g^{\mu\nu}\right).
\label{leptonic}
\eeq
In order to simplify the discussion, we will mainly consider the metric part $L^{\mu\nu}\to -Q^2g^{\mu\nu}$.  We will discuss the result with the full leptonic tensor (\ref{leptonic}) in the end of Sec. III.
The SSA in SIDIS can be generated by both the initial state and final state twist-3 contributions. 
In this paper, we focus on the contribution from initial state twist-3 distribution functions of the
transversely polarized proton, then the polarized differential cross section can be written as
\beq
{d^4\Delta\sigma\over dx_Bdydz_hdP_{h\perp}}={\alpha^2_{em}\over 32\pi^2z_hx_B^2S^2_{ep}Q^2}\sum_i\int{{dz\over z^2}} W_i D_{i\to h}(z),
\eeq
where $D_{i\to h}(z)$ is the twist-2 unpolarized fragmentation function.
The hadronic part $W_i$ describes a scattering of the virtual photon on the transversely polarized proton, with the leptonic metric part contracted. We will make the subscript $i$ implicit in the rest part of this paper for simplicity.

%%%%%%%%%%%%%%%%%%%%%%%%%%%%%%%%%%%%%%%%%%%%%%%%%%%%%%%%%%%%%%%%%%%%%%%%%%%%%%%
\begin{figure}[h]
\begin{center}
  \includegraphics[height=4cm,width=12cm]{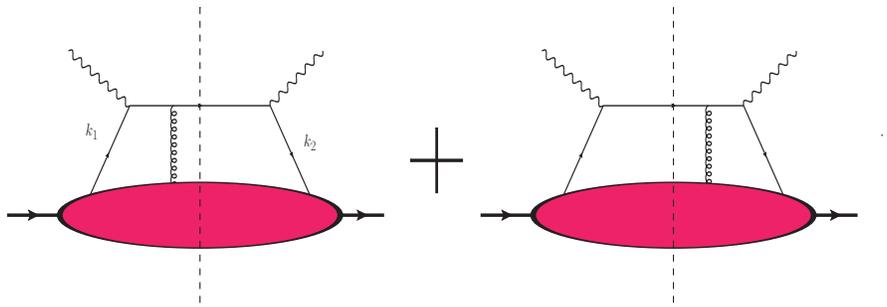}\hspace{1cm}
\end{center}
 \caption{Diagrammatic description of Eq. (\ref{diagram_method_pole}).}
\label{fig1}
\end{figure}
%%%%%%%%%%%%%%%%%%%%%%%%%%%%%%%%%%%%%%%%%%%%%%%%%%%%%%%%%%%%%%%%%%%%%%%%%%%%%%% 

In the conventional pole calculation, one needs to consider diagrams as shown in Fig. \ref{fig1}, in which the hadronic part reads
\beq
W_{\rm pole}=\int{d^4k_1\over (2\pi)^4}\int{d^4k_2\over (2\pi)^4}
\int d^4 y_1 \int d^4y_2\,e^{ik_1\cdot y_1}e^{i(k_2-k_1)\cdot y_2}
\la pS_{\perp}|\bar{\psi}_j(0)gA^{\alpha}(y_2)\psi_i(y_1)|pS_{\perp}\ra
H^{{\rm pole}}_{ji,\alpha}(k_1,k_2).
\label{diagram_method_pole}
\eeq
The twist-3 contribution is generated by the pole term, which comes from the imaginary part of the quark propagator
\beq
{1\over k^2+i\epsilon}=P\left(1\over k^2\right)-i\pi\delta(k^2),
\label{separation}
\eeq
thus only $-i\pi\delta(k^2)$ is considered in the conventional pole method. 
We perform collinear expansion $k_i\to x_ip$ for the hard part $H^{{\rm pole}}_{ji,\alpha}(k_1,k_2)$.
There are three types of pole contributions at the leading-order with respect to QCD coupling constant $\alpha_s$.
: soft-gluon-pole(SGP, $x_2-x_1=0$),
soft-fermion-pole(SFP, $x_1=0$ or $x_2=0$, $x_1\neq x_2$) and hard-pole(HP, $x_1=x_B,~x_2\neq x_B$ or $x_2=x_B, x_1\neq x_B$) contributions. 
Full diagrams for each pole contribution are shown in Fig. 2a-2c.
%%%%%%%%%%%%%%%%%%%%%%%%%%%%%%%%%%%%%%%%%%%%%%%%%%%%%%%%%%%%%%%%%%%%%%%%%%%%%%%
\setcounter{section}{1}
\renewcommand{\thefigure}{\arabic{figure}\alph{section}}
\begin{figure}[h]
\begin{center}
  \includegraphics[height=8cm,width=12cm]{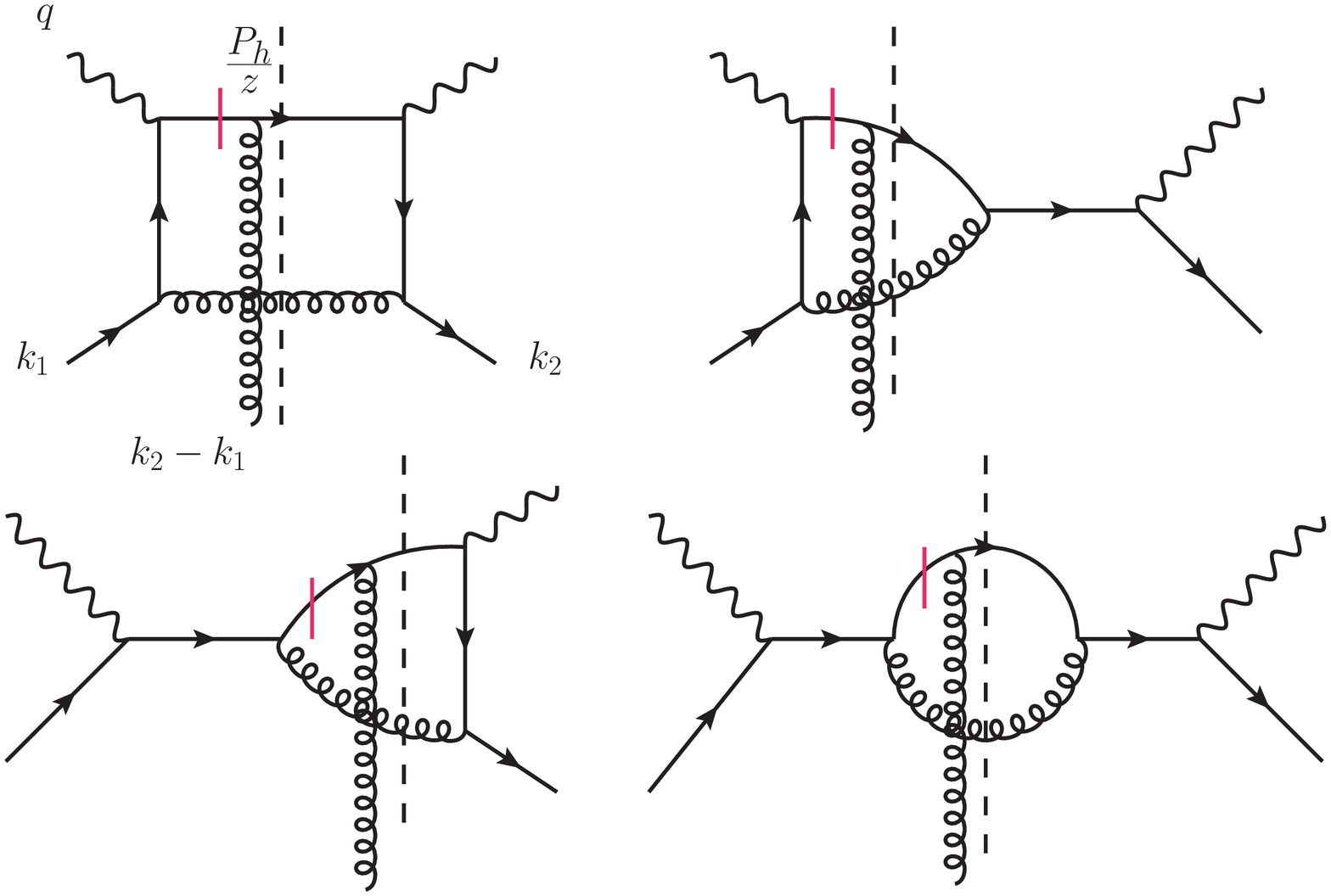}\hspace{1cm}
  \end{center}
  \caption{The diagrams for SGP contribution. The separation of the pole (\ref{separation}) is carried out for the red barred propagators. The complex conjugate diagrams also need to be considered.}
  \label{fig2(a)}
\end{figure}
%%%%%%%%%%%
\setcounter{figure}{1}
\setcounter{section}{2}
\begin{figure}[h]
\begin{center}
\includegraphics[height=4cm,width=12cm]{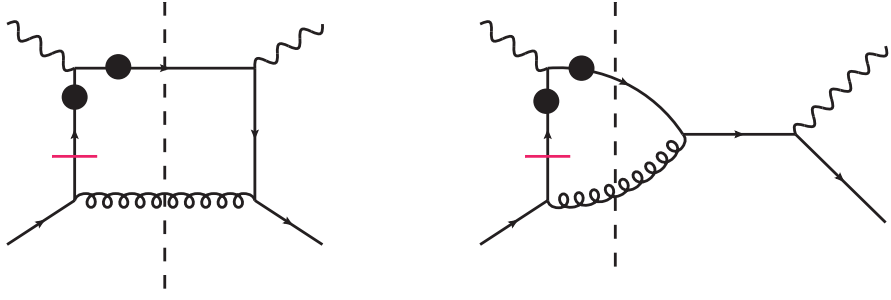}\hspace{1cm}
\end{center}
\caption{The diagrams for SFP contribution. The gluon line with momentum $k_2-k_1$ attaches to each black dot.}
  \label{fig2(b)}
\end{figure}
%%%%%%%%%%%%
\setcounter{figure}{1}
\setcounter{section}{3}
\begin{figure}[h]
\begin{center}
  \includegraphics[height=4cm,width=12cm]{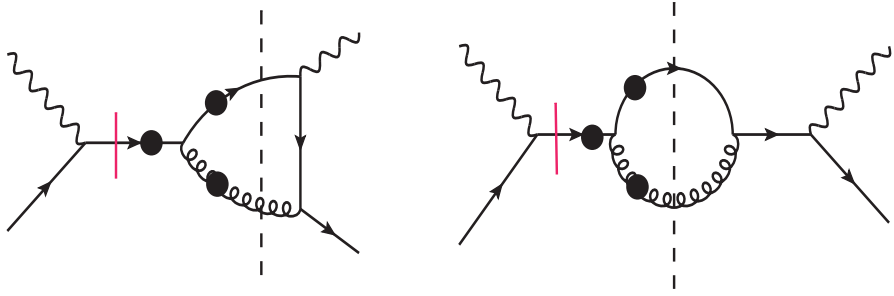}\hspace{1cm}
\end{center}
 \caption{The diagrams for HP contribution.}
\label{fig2(c)}
\end{figure}
%%%%%%%%%%%%%%%%%%%%%%%%%%%%%%%%%%%%%%%%%%%%%%%%%%%%%%%%%%%%%%%%%%%%%%%%%%%%%%% 
It's known that there is other type of contribution given by diagrams with two quark lines in the same side of the cut~\cite{Yoshida:2016tfh}. This contribution is relatively easier to calculate, thus will not be discussed in this paper. 

One can factor out the $\delta$-functions in $H^{{\rm pole}}_{ji,\alpha}(k_1,k_2)$ for the three pole contributions
\beq
H^{{\rm pole}}_{ji,\alpha}(k_1,k_2)
&=&H^{{\rm SGP}}_{Lji,\alpha}(k_1,k_2)\Bigl\{-i\pi\delta\Bigl[({p_c}-(k_2-k_1))^2\Bigr]\Bigr\}
(2\pi)\delta\Bigl[(k_2+q-{p_c})^2\Bigr]
\nonumber\\
&&+H^{{\rm SFP}}_{Lji,\alpha}(k_1,k_2)\Bigl\{-i\pi\delta\Bigl[(p_c-k_2+k_1-q)^2\Bigr]\Bigr\}
(2\pi)\delta\Bigl[(k_2+q-{p_c})^2\Bigr]
\nonumber\\
&&+H^{{\rm HP}}_{Lji,\alpha}(k_1,k_2)\Bigl\{-i\pi\delta\Bigl[(k_1+q)^2\Bigr]\Bigr\}
(2\pi)\delta\Bigl[(k_2+q-{p_c})^2\Bigr]
\nonumber\\
&&+({\rm complex\ conjugate\ diagrams}),
\eeq
where the factor $(2\pi)\delta\Bigl[(k_2+q-{p_c})^2\Bigr]$ representing the on-shell condition of the unobserved
parton, $p_c$ is the four-momentum of the final state fragmenting parton. 

A systematic way to calculate the pole contributions was developed in~\cite{Eguchi:2006mc}. 
We confirmed that Ward-Takahashi identity(WTI) shown in~\cite{Eguchi:2006mc} is valid for the diagrams in Fig. 2a-2c as
\beq
(k_2-k_1)^{\alpha}H^{{\rm pole}}_{ji,\alpha}(k_1,k_2)=0.
\label{WTI_pole}
\eeq
Considering $k_1$ and $k_2$ derivatives, we can derive relations,
\beq
(x_2-x_1){\partial\over \partial k_1^{\beta}}H^{{\rm pole}}_{ji, p}(k_1,k_2)\Bigl|_{k_i=x_ip}
&=&H^{{\rm pole}}_{ji,\beta}(x_1p,x_2p),
\nonumber\\
(x_2-x_1){\partial\over \partial k_2^{\beta}}H^{{\rm pole}}_{ji, p}(k_1,k_2)\Bigl|_{k_i=x_ip}
&=&-H^{{\rm pole}}_{ji, \beta}(x_1p,x_2p),
\label{WTI_pole1}
\eeq
where $H^{{\rm pole}}_{ji, p}(k_1,k_2) = p^{\beta}H^{{\rm pole}}_{ji, \beta}(k_1,k_2)$.
Thus we can derive the following useful relations for SFP and HP
\beq
{\partial\over \partial k_1^{\beta}}H^{{\rm SFP(HP)}}_{ji,p}(k_1,k_2)\Bigl|_{k_i=x_ip}
&=&{1\over x_2-x_1}H^{{\rm SFP(HP)}}_{ji,\beta}(x_1p,x_2p),
\nonumber\\
{\partial\over \partial k_2^{\beta}}H^{{\rm SFP(HP)}}_{ji,p}(k_1,k_2)\Bigl|_{k_i=x_ip}
&=&-{1\over x_2-x_1}H^{{\rm SFP(HP)}}_{ji,\beta}(x_1p,x_2p),
\label{WTI_pole2}
\eeq
which lead to
\beq
-{\partial\over \partial k_1^{\beta}}H^{{\rm SFP(HP)}}_{ji,p}(k_1,k_2)\Bigl|_{k_i=x_ip}
={\partial\over \partial k_2^{\beta}}H^{{\rm SFP(HP)}}_{ji,p}(k_1,k_2)\Bigl|_{k_i=x_ip}.
\label{k1eqk2}
\eeq
However, we cannot derive the same relation with Eqs. (\ref{WTI_pole2},\ref{k1eqk2}) for the SGP diagrams 
directly from WTI because $H^{{\rm SGP}}_{Lji,\alpha}(x_1p,x_2p)$ contains $\delta(x_2-x_1)$.
So far, the only way to derive the relations is to calculate all the relevant diagrams explicitly, which is annoying in higher order perturbative QCD calculations. In SIDIS at $\mathcal{O}(\alpha_s)$, the authors of Ref.~\cite{Eguchi:2006mc} have checked explicitly that the above relation also holds true
for $H^{{\rm SGP}}_{ji,p}$,
\beq
-{\partial\over \partial k_1^{\beta}}H^{{\rm SGP}}_{ji,p}(k_1,k_2)\Bigl|_{k_i=x_ip}
={\partial\over \partial k_2^{\beta}}H^{{\rm SGP}}_{ji,p}(k_1,k_2)\Bigl|_{k_i=x_ip}.
\label{relationSGP}
\eeq
Now we can perform collinear expansion of the hard part
\beq
&&H^{{\rm pole}}_{ji,\rho}(k_1,k_2)
\nonumber\\
&\simeq& H^{{\rm pole}}_{ji,\rho}(x_1 p, x_2 p)
+{\partial\over \partial k_1^{\alpha}}H^{{\rm pole}}_{ji,\rho}(k_1,k_2)\Bigl|_{k_i=x_i p}
\omega^{\alpha}_{\ \beta}k_1^{\beta}
+{\partial\over \partial k_2^{\alpha}}H^{{\rm pole}}_{ji,\rho}(k_1,k_2)\Bigl|_{k_i=x_i p}
\omega^{\alpha}_{\ \beta}k_2^{\beta}
\nonumber\\
&=&H^{{\rm pole}}_{ji,\rho}(x_1 p,x_2 p)
+{\partial\over \partial k_2^{\alpha}}H^{{\rm pole}}_{ji,\rho}(k_1,k_2)\Bigl|_{k_i=x_i p}
\omega^{\alpha}_{\ \beta}(k_2-k_1)^{\beta},
\label{collinear}
\eeq
where the projection tensor is defined as $\omega^{\alpha}_{\ \beta}=g^{\alpha}_{\ \beta}-\bar n^{\alpha}n_{\beta}$ with the unit vectors taken as $\bar n = [1,0,0],~n=[0,1,0]$. We work in the hadron frame and $p^{\mu} = p^+\bar n^{\mu}$.
We can neglect $k^{-}$-component at the twist-3 accuracy and identify $\omega^{\alpha}_{\ \beta}k_{i}^{\beta}\simeq k^{\alpha}_{i\perp}$. We use the same identification for all vectors projected by
$\omega^{\alpha}_{\ \beta}$ below.
The next step is to decompose the components of the gluon field $A^{\alpha}$ into longitudinal and transverse as
\beq
A^{\alpha}\simeq \frac{A^{n}}{p^+}p^{\alpha}+A^{\alpha}_{\perp},
\label{decomposition}
\eeq
where $A^n=A\cdot n$ and we neglected $A^-$-component.
Substitute Eqs. (\ref{collinear},\ref{decomposition}) into Eq. (\ref{diagram_method_pole}), 
we can extract the twist-3 contribution
\beq
W_{\rm pole}&=&\int{d^4k_1\over (2\pi)^4}\int{d^4k_2\over (2\pi)^4}\int d^4y_1\int d^4y_2
\,e^{ik_1\cdot y_1}e^{i(k_2-k_1)\cdot y_2}
\la pS_{\perp}|\bar{\psi}_j(0)gA^{n}(y_2)\psi_i(y_1)|pS_{\perp}\ra
\nonumber\\
&&\times\frac{1}{p^+}{\partial\over \partial k_2^{\alpha}}H^{{\rm pole}}_{ji,p}(k_1,k_2)\Bigl|_{k_i=(k_i\cdot n)p}
(k_{2\perp}-k_{1\perp})^{\alpha}
\nonumber\\
&+&\int{d^4k_1\over (2\pi)^4}\int{d^4k_2\over (2\pi)^4}
\int d^4y_1 \int d^4y_2\,e^{ik_1\cdot y_1}e^{i(k_2-k_1)\cdot y_2}
\la pS_{\perp}|\bar{\psi}_j(0)gA_{\perp}^{\alpha}(y_2)\psi_i(y_1)|pS_{\perp}\ra
\nonumber\\
&&\times H^{{\rm pole}}_{ji,\alpha}(x_1p,x_2p)
\nonumber\\
&=&ip^+ \int{dx_1}\int{dx_2}\int{dy_1^-\over 2\pi}\int{dy_2^-\over 2\pi}
\,e^{i x_1p^+y_1^-}e^{i(x_2-x_1)p^+y_2^-}
\la pS_{\perp}|\bar{\psi}_j(0)g\Bigl[\partial_{\perp}^{\alpha}A^{n}(y_2^-)-\partial^nA_{\perp}^{\alpha}(y_2^-)\Bigr]\psi_i(y_1^-)|pS_{\perp}\ra
\nonumber\\
&&\times{\partial\over \partial k_2^{\alpha}}H^{{\rm pole}}_{ji,p}(k_1,k_2)\Bigl|_{k_i=x_ip}
\nonumber\\
&+&(p^+)^2\int{dx_1}\int{dx_2}\int{dy_1^-\over 2\pi}\int{dy_2^-\over 2\pi}
\,e^{ix_1p^+y_1^-}e^{i(x_2-x_1)p^+y_2^-}
\la pS_{\perp}|\bar{\psi}_j(0)gA_{\perp}^{\alpha}(y_2^-)\psi_i(y_1^-)|pS_{\perp}\ra
\nonumber\\
&&\times \Bigl[(x_2-x_1){\partial\over \partial k_2^{\alpha}}H^{{\rm pole}}_{ji,p}(k_1,k_2)\Bigl|_{k_i=x_ip}
+H^{{\rm pole}}_{ji,\alpha}(x_1p,x_2p)\Bigr].
\label{twist-3_pole}
\eeq
The last term in Eq. (\ref{twist-3_pole}) can be eliminated by using the relation Eq. (\ref{WTI_pole1}), and the first term can be rewritten as
\beq
W_{\rm pole}=i\int{dx_1}\int{dx_2} {\rm Tr}\left[M_F^{\alpha}(x_1,x_2)
{\partial\over \partial k_{2}^{\alpha}}H^{{\rm pole}}_{p}(k_1,k_2)\Bigl|_{k_i=x_ip}\right],
\label{twist-3_pole_2}
\eeq
where $M_F^{\alpha}(x_1,x_2)$ is the F-type dynamical function which can be further expanded as
\beq
M_{ij, F}^{\alpha}(x_1,x_2)&=&p^+ \int{dy_1^-\over 2\pi}\int{dy_2^-\over 2\pi}
\,e^{i x_1p^+y_1^-}e^{i(x_2-x_1)p^+y_2^-}
\la pS_{\perp}|\bar{\psi}_j(0)gF^{\alpha n}(y_2^-)\psi_i(y_1^-)|pS_{\perp}\ra 
\nonumber\\
&=&-{M_N\over 2}\epsilon^{\alpha \bar n nS_{\perp}}(\slash{p})_{ij}T_{q,F}(x_1,x_2)+\cdots,
\label{eq-MF_def}
\eeq
with the nucleon mass $M_N$ and the field strength tensor defined as $F^{\alpha n}(y_2^-) = \partial_{\perp}^{\alpha}A^{n}(y_2^-)-\partial^nA_{\perp}^{\alpha}(y_2^-)$, notice that the nonlinear gluon term in the field strength tensor has been omitted because it comes from Feynman diagrams with linked gluons more than one, therefore does not show in Eq. (\ref{twist-3_pole_2}). $T_{q,F}(x_1,x_2)$ is the well-known Qiu-Sterman function, defined as following\footnote{We rescaled the function as $T_{q,F}(x_1,x_2)\to (g/2\pi M_N)T_{q,F}(x_1,x_2)$ 
from the original definition in Ref.~\cite{Qiu:1998ia}
for convenience. Our definition of $T_{q,F}(x_1,x_2)$ is the same with $F^q_{FT}(x_2,x_1)$ in Ref.~\cite{Kanazawa:2015ajw}.}
\bea
T_{q,F}(x_1,x_2) =\Bigl({g\over 2\pi M_N}\Bigr)\int{dy_1^- dy_2^-\over 4\pi}
\,e^{i x_1p^+y_1^-}e^{i(x_2-x_1)p^+y_2^-}
\la pS_{\perp}|\bar{\psi}(0)\slash{n}\epsilon^{\alpha \bar n nS_{\perp}}F_{\alpha}^{\ n}(y_2^-)\psi(y_1^-)|pS_{\perp}\ra.
\label{eq-MF_def2}
\eea
Using Eq. (\ref{WTI_pole2}) for SFP and HP, one can evaluate the derivative of the hard part with respect to $k_2$ in Eq. (\ref{twist-3_pole_2}). For SGP, we need to rely on the master formula \cite{Koike:2006qv}
\beq
{\partial\over \partial k_2^{\alpha}}H^{{\rm SGP}}_{ji,p}(k_1,k_2)\Bigl|_{k_i=x_ip}
={1\over 2NC_F}\Bigl[i\pi\delta(x_2-x_1)\Bigr]\left({\partial\over \partial p_c^{\alpha}}
-{p_{c\alpha}p^{\mu}\over p_c\cdot p}{\partial\over \partial p_c^{\mu}} \right)H_{ji}(x_1p),
\hspace{5mm}
\label{master}
\eeq
where $H_{ji}(x_1p)$ is the usual $2\to 2$ $\gamma^*q\to qg$ scattering cross section without the extra gluon line attached.
%where we introduced the term
%\beq
%&&-i\omega^{\alpha}_{\ \beta}\int{dx}\int{dx'}\int{d\lambda\over 2\pi}\int{d\mu\over 2\pi}
%\,e^{i\lambda x'}e^{i\mu(x-x')}
%\la pS_{\perp}|\bar{\psi}_j(0)g\partial^nA^{\beta}(\mu n)\psi_i(\lambda n)|pS_{\perp}\ra
%{\partial\over \partial k_2^{\alpha}}H^{{\rm pole}\,\mu\nu}_{ji\,p}(k_1,k_2)\Bigl|_{k_i=x_ip}
%\nonumber\\
%&=&-\omega^{\alpha}_{\ \beta}\int{dx}\int{dx'}\int{d\lambda\over 2\pi}\int{d\mu\over 2\pi}
%\,e^{i\lambda x'}e^{i\mu(x-x')}
%\la pS_{\perp}|\bar{\psi}_j(0)gA^{\beta}(\mu n)\psi_i(\lambda n)|pS_{\perp}\ra
%(x-x'){\partial\over \partial k_2^{\alpha}}H^{{\rm pole}\,\mu\nu}_{ji\,\rho}(k_1,k_2)\Bigl|_{k_i=x_ip}.
%\hspace{6mm}
%\eeq
Combining the three pole contributions together, we obtain the final result based on the conventional pole calculation 
\beq
&&{d^4\Delta\sigma\over dx_Bdydz_hdP_{h\perp}}
\nonumber\\
&=&{\pi M_N\alpha^2_{em}\alpha_s\over 8z_hx_B^2S^2_{ep}Q^2}\sum_qe^2_q\int{{dz\over z^2}}D_{q\to h}(z)\int{dx\over x}
\delta\Bigl[(xp+q-p_c)^2\Bigr]\Bigl((\hat{s}+Q^2)\epsilon^{p_c\bar{n}nS_{\perp}}+\hat{t}\epsilon^{q\bar{n}nS_{\perp}}\Bigr)
\nonumber\\
&&\times\Bigl[x{d\over dx}T_{q,F}(x,x)\hat{\sigma}_{D}+T_{q,F}(x,x)\hat{\sigma}_{ND}
+T_{q,F}(0,x)\hat{\sigma}_{SFP}
+T_{q,F}(x_B,x)\hat{\sigma}_{HP}
\Bigr],
\label{result_pole}
\eeq
where all hard cross sections are listed below,
\beq
\hat{\sigma}_{D}&=&{1\over 2N}{16Q^2\left[(\hat{s}+\hat{t})^2+(\hat{t}+\hat{u})^2\right]\over \hat{s}\hat{t}\hat{u}^2},
\nonumber\\
\hat{\sigma}_{ND}&=&{1\over 2N}{16Q^2[2Q^6+\hat{t}^3-4Q^2\hat{s}\hat{u}+3\hat{t}^2\hat{u}+4\hat{t}\hat{u}^2
+2\hat{u}^3+Q^4(3\hat{t}+\hat{u})]\over \hat{s}^2\hat{t}\hat{u}^2},
\nonumber\\
\hat{\sigma}_{SFP}&=&-{1\over 2N}{16Q^2[2Q^4+\hat{u}^2+Q^2(\hat{t}+3\hat{u})]\over \hat{s}\hat{t}\hat{u}^2},
\nonumber\\
\hat{\sigma}_{HP}&=&\Bigl({1\over 2N}{1\over \hat{t}}-C_F{1\over \hat{s}+Q^2}\Bigr)
{16Q^2[Q^6+3Q^4\hat{s}+\hat{s}^3+Q^2(3\hat{s}^2+\hat{t}^2)]\over \hat{s}^2\hat{u}^2},
\label{eq-sigmahat}
\eeq
with the standard Mandelstam variables defined as
\beq
\hat{s}&=&(xp+q)^2,\hspace{5mm}
\hat{t}=(xp-p_c)^2,\hspace{5mm}
\hat{u}=(q-p_c)^2.
\eeq
In the next section, we show that the new non-pole method can reproduce these hard cross sections.
We would like to make a comment on the relation Eq. (\ref{relationSGP}) for the SGP diagrams,
this relation is required to construct the gauge-invariant matrix for the dynamical function. 
However, there is no simple way to prove this relation. We have to check if it's correct diagram by diagram.
This is a frustrating point of the conventional pole calculation. We will show that
the new method can avoid such complexity and thus a more flexible calculation technique.

%-------------------   Section 3   --------------

\setcounter{section}{2}
\section{The new non-pole calculation for Sivers effect}
We introduce the new non-pole calculation method in this section. The main difference between 
the pole and the non-pole methods is on the decomposition of the propagator shown in 
Eq. (\ref{separation}). 
In the new method, we directly perform the contour integrations and never carry out  
the decomposition for any propagators.
The new method is expected to remove the mathematical complexity lies in the validity of the
decomposition.
In this sense, the new method can be regarded as a more flexible approach, and can be easily
extended to more complicated cases. Removal of the workload on Eq.~(\ref{relationSGP}) is one of 
important consequences.

\subsection{General formalism}
%%%%%%%%%%%%%%%%%%%%%%%%%%%%%%%%%%%%%%%%%%%%%%%%%%%%%%%%%%%%%%%%%%%%%%%%%%%%%%%
\setcounter{figure}{2}
\renewcommand{\thefigure}{\arabic{figure}}
\begin{figure}[h]
\begin{center}
  \includegraphics[height=4cm,width=6cm]{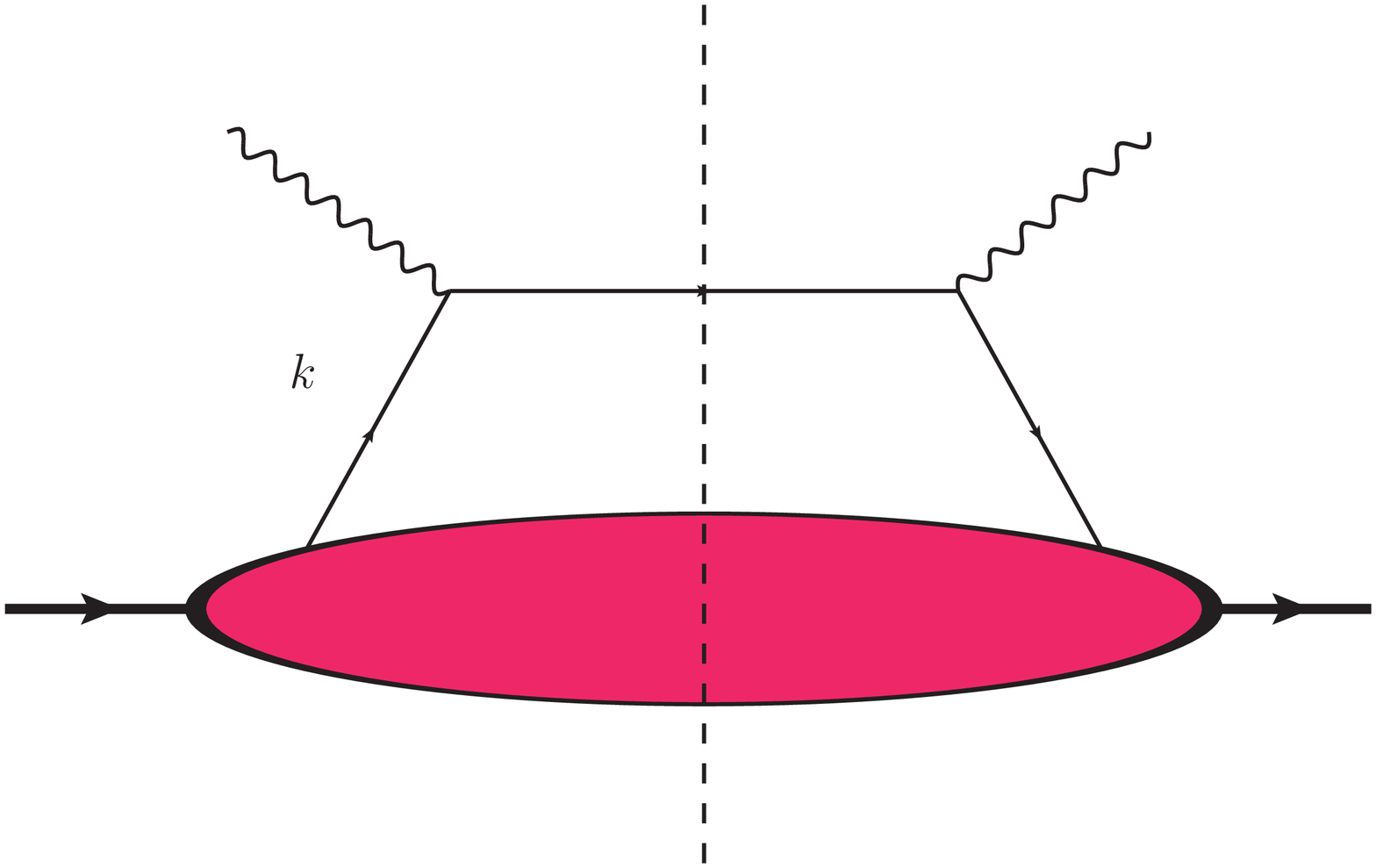}\hspace{1cm}
\end{center}
 \caption{Diagrammatic description of Eq. (\ref{diagram1}).}
\label{fig3}
\end{figure}
%%%%%%%%%%%%%%%%%%%%%%%%%%%%%%%%%%%%%%%%%%%%%%%%%%%%%%%%%%%%%%%%%%%%%%%%%%%%%%% 
In the new method we propose here, the hadronic part should be written as a sum of all the diagrams, i.e., $W = \sum_i W^{(i)}$, where $i$ denotes the number of gluon attachment. Let's start with the diagram in Fig. \ref{fig3} without any gluon attached, the hadronic part is given by
\beq
W^{(0)} = \int{d^4k\over (2\pi)^4}\int d^4y_1\,e^{ik\cdot y_1}\la pS_{\perp}|\bar{\psi}_j(0)\psi_i(y_1)|pS_{\perp}\ra H_{ji}(k).
\label{diagram1}
\eeq
The twist-3 contribution from the diagrams without gluon attached can be obtained by performing collinear expansion of the hard part
\beq
H_{ji}(k)\simeq H_{ji}(xp)+{\partial\over \partial k^{\alpha}}H_{ji}(k)\Bigl|_{k=xp}
k_{\perp}^{\alpha}.
\eeq
Then Eq. (\ref{diagram1}) can be expanded into two parts
\beq
W^{(0)} &=&\int{d^4k\over (2\pi)^4}\int d^4y_1\,e^{ik\cdot y_1}\la pS_{\perp}|\bar{\psi}_j(0)\psi_i(y_1)|pS_{\perp}\ra
\Biggl[H_{ji}(xp)+{\partial\over \partial k^{\alpha}}H_{ji}(k)\Bigl|_{k=xp}k_{\perp}^{\alpha}\Biggr]
\nonumber\\
&=& p^+\int{dx}\int{dy_1^- \over 2\pi}\,e^{i xp^+y_1^-}\la pS_{\perp}|\bar{\psi}_j(0)\psi_i(y_1^-)|pS_{\perp}\ra H_{ji}(xp)
\nonumber\\
&&+ip^+ \int{dx}\int{dy_1^- \over 2\pi}\,e^{i xp^+ y_1^-}
\la pS_{\perp}|\bar{\psi}_j(0)\partial_{\perp}^{\alpha}\psi_i(y_1^-)|pS_{\perp}\ra
{\partial\over \partial k^{\alpha}}H_{ji}(k)\Bigl|_{k=xp}.
\label{eq-W0}
\eeq
In general, the first term could give the twist-3 contribution when the hard part gives
transverse component $H_{ji}(xp)\sim (\gamma^{\perp})_{ji}$. Next we consider the diagrams with one gluon attachment as shown in Fig. \ref{fig1} which were also considered in the conventional method. 
Here we need to consider a set of diagrams $H_{ji,\alpha}(k_1,k_2)$ shown in Fig. \ref{fig4} and their complex conjugate.
%%%%%%%%%%%%%%%%%%%%%%%%%%%%%%%%%%%%%%%%%%%%%%%%%%%%%%%%%%%%%%%%%%%%%%%%%%%%%%%
\begin{figure}[h]
\begin{center}
  \includegraphics[height=8cm,width=12cm]{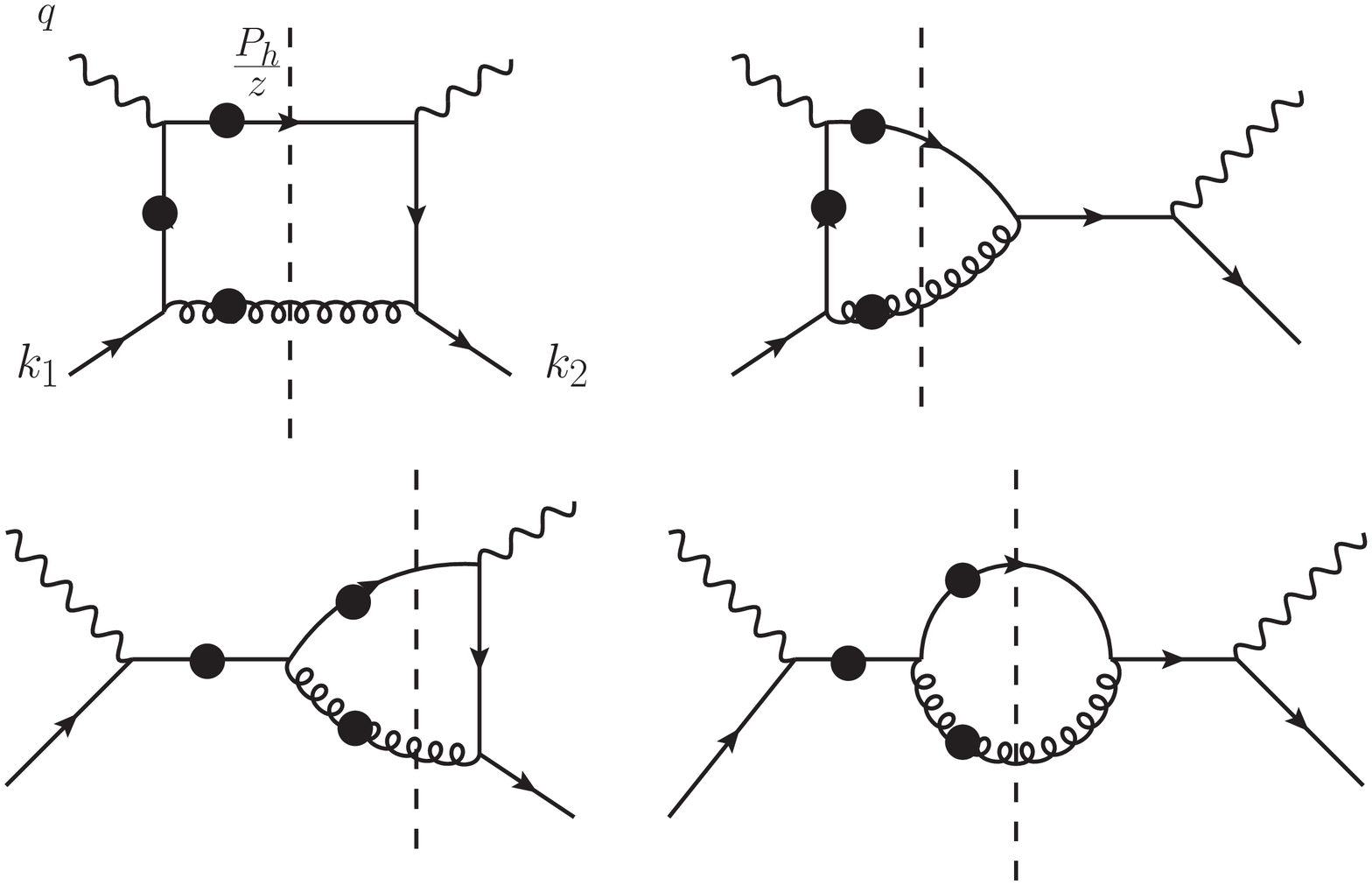}\hspace{1cm}
\end{center}
 \caption{The diagrams for $H_{ji,\alpha}(k_1,k_2)$. The gluon line with momentum $k_2-k_1$ attaches to each black dot.}
\label{fig4}
\end{figure}
%%%%%%%%%%%%%%%%%%%%%%%%%%%%%%%%%%%%%%%%%%%%%%%%%%%%%%%%%%%%%%%%%%%%%%%%%%%%%%% 
We call them non-pole diagrams because we don't separate the pole term for any propagators. The non-pole contribution to the hadronic part reads
\beq
W^{(1)} = \int{d^4k_1\over (2\pi)^4}\int{d^4k_2\over (2\pi)^4}\int d^4y_1\int d^4y_2\,
e^{ik_1\cdot y_1}e^{i(k_2-k_1)\cdot y_2}\la pS_{\perp}|\bar{\psi}_j(0)gA^{\alpha}(y_2)\psi_i(y_1)|pS_{\perp}\ra H_{ji,\alpha}(k_1,k_2).
\label{eq-W1_1}
\eeq
Similar to the strategy in dealing with the diagrams without gluon attachment, the first step to extract the twist-3 contribution from one gluon attached diagrams is to perform collinear expansion of the hard part
\bea
H_{ji,\rho}(k_1,k_2) = H_{ji,\rho}(x_1p, x_2p) +
{\partial H_{ji,\rho}(k_1,k_2) \over \partial k_1^{\alpha}}\Bigl|_{k_i=x_ip}k_{1\perp}^{\alpha}
+{\partial H_{ji,\rho}(k_1,k_2) \over \partial k_2^{\alpha}}\Bigl|_{k_i=x_ip}k_{2\perp}^{\alpha}.
\label{eq-H1_exp}
\eea
One also need to decompose the gluon field $A^{\alpha}$ into longitudinal and transverse components
as in Eq. (\ref{decomposition}). Then Eq.(\ref{eq-W1_1}) can be expanded as follows:
\bea
W^{(1)}
=&\int{d^4k_1\over (2\pi)^4}\int{d^4k_2\over (2\pi)^4}\int d^4y_1\int d^4y_2
\,e^{ik_1\cdot y_1}e^{i(k_2-k_1)\cdot y_2}
\la pS_{\perp}|\bar{\psi}_j(0)gA^{n}(y_2)\psi_i(y_1)|pS_{\perp}\ra
\nonumber\\
&\times{1\over p^+}\Bigl[H_{ji,p}(x_1p,x_2p)
+{\partial\over \partial k_1^{\alpha}}H_{ji,p}(k_1,k_2)\Bigr|_{k_i=x_ip}
k_{1\perp}^{\alpha}+{\partial\over \partial k_2^{\alpha}}H_{ji,p}(k_1,k_2)
\Bigr|_{k_i=x_ip}k_{2\perp}^{\alpha}\Bigr]
\nonumber\\
&+\int{d^4k_1\over (2\pi)^4}\int{d^4k_2\over (2\pi)^4}
\int d^4y_1\int d^4y_2\,e^{ik_1\cdot y_1}e^{i(k_2-k_1)\cdot y_2}
\la pS_{\perp}|\bar{\psi}_j(0)gA_{\perp}^{\alpha}(y_2)\psi_i(y_1)|pS_{\perp}\ra
H_{ji,\alpha}(x_1p,x_2p),
\label{eq-W1_2}
\eea
notice that other terms in the combination of Eqs. (\ref{decomposition}) and (\ref{eq-H1_exp}) contribute 
to higher twist. The hard part shown in above equation can be further simplified by using the WTI relations. It is straightforward to derive the counterpart of Eq.~(\ref{WTI_pole}) for the non-pole diagrams
\cite{Kanazawa:2013uia}
\beq
(k_2-k_1)^{\alpha}H_{ji,\alpha}(k_1,k_2)=H_{ji}(k_2)-H_{ji}(k_1).
\label{WTI-nonpole}
\eeq
The non-pole hard part doesn't have the delta function $\delta(x_2-x_1)$, therefore, we can derive the following useful relations.
\beq
H_{ji,p}(x_1p,x_2p)
&=&{1\over x_2-x_1-i\epsilon}\Bigl[H_{ji}(x_2p)-H_{ji}(x_1p)\Bigr],
\nonumber\\
{\partial\over \partial k_1^{\beta}}H_{ji,p}(k_1,k_2)\Bigr|_{k_i=x_ip}
&=&{1\over x_2-x_1-i\epsilon}\Bigl[ H_{ji,\beta}(x_1p,x_2p)
-{\partial\over \partial k_1^{\beta}}H_{ji}(k_1)\Bigr|_{k_1=x_1p}\Bigr],
\label{WTI-nonpole2}\\
{\partial\over \partial k_2^{\beta}}H_{ji,p}(k_1,k_2)\Bigr|_{k_i=x_ip}
&=&-{1\over x_2-x_1-i\epsilon}\Bigl[H_{ji,\beta}(x_1p,x_2p)
-{\partial\over \partial k_2^{\beta}}H_{ji}(k_2)\Bigr|_{k_2=x_2p}\Bigr],
\nonumber
\eeq
where the sign of $i\epsilon$ was determined by the fact that only the final state interaction exists in SIDIS. If a process has both the initial and the final interactions as in the case of $pp$-collision, 
the r.h.s of Eq.~(\ref{WTI-nonpole2}) could have both $\pm i\epsilon$ and the sign can't be uniquely determined from the WTI (\ref{WTI-nonpole}). We need more consideration on this point when we apply the same technique to $pp$-collision. 

We would like to emphasize the validity of WTI to higher order diagrams. The WTI is 
a consequence of the gauge invariance in QCD and therefore we can use the same 
Eqs. (\ref{WTI-nonpole},\ref{WTI-nonpole2}) to higher order diagrams as long as we use 
an appropriate regularization scheme like the dimensional regularization.
By using these useful relations derived from WTI, the hard part terms $H_{ji,\alpha}(k_1,k_2)$ and $H_{ji,p}(k_1,k_2)$ contained in Eq. (\ref{eq-W1_2}) are respectively given by
\bea
&H_{ji,\alpha}(x_1 p, x_2 p)
=-(x_2-x_1){\partial\over \partial k_2^{\alpha}}H_{ji,p}(k_1,k_2)\Bigr|_{k_i=x_ip}
+{\partial\over \partial k_2^{\alpha}}H_{ji}(k_2)\Bigr|_{k_2=x_2p}
\nonumber\\
&\hspace{23mm}=(x_2-x_1)\Biggl[{1\over x_2-x_1-i\epsilon}
H_{ji,\alpha}(x_1p,x_2p)-{1\over x_2-x_1-i\epsilon}
{\partial\over \partial k_2^{\alpha}}H_{ji}(k_2)\Bigr|_{k_2=x_2p}\Biggr]
+{\partial\over \partial k_2^{\alpha}}H_{ji}(k_2)\Bigr|_{k_2=x_2p},
\nonumber\\
&H_{ji,p}(x_1p,x_2p)
+{\partial\over \partial k_2^{\alpha}}H_{ji,p}(k_1,k_2)
\Bigr|_{k_i=x_ip}k_{2\perp}^{\alpha}
+{\partial\over \partial k_1^{\alpha}}H_{ji,p}(k_1,k_2)\Bigr|_{k_i=x_ip}
k_{1\perp}^{\alpha}
\nonumber\\
=&{1\over x_2-x_1-i\epsilon}\Bigl[H_{ji}(x_2p)-H_{ji}(x_1p)\Bigr]
+{\partial\over \partial k_2^{\alpha}}H_{ji,p}(k_1,k_2)
\Bigr|_{k_i=x_ip}(k_{2\perp}-k_{1\perp})^{\alpha}
\nonumber\\
&+\Bigl({\partial\over \partial k_1^{\alpha}}H_{ji,p}(k_1,k_2)\Bigr|_{k_i=x_ip}+{\partial\over \partial k_2^{\alpha}}H_{ji,p}(k_1,k_2)\Bigr|_{k_i=x_ip}\Bigr)
k_{1\perp}^{\alpha}
\nonumber\\
=&{1\over x_2-x_1-i\epsilon}\Biggl\{\Bigl[H_{ji}(x_2p)-H_{ji}(x_1p)\Bigr]
-H_{ji,\alpha}(x_1p,x_2p)(k_{2\perp}-k_{1\perp})^{\alpha}
+{\partial\over \partial k_2^{\alpha}}H_{ji}(k_2)\Bigr|_{k_2=x_2p}
(k_{2\perp}-k_{1\perp})^{\alpha}
\nonumber\\
&+\Biggl[
{\partial\over \partial k_2^{\alpha}}H_{ji}(k_2)\Bigr|_{k_2=x_2p}
-{\partial\over \partial k_1^{\alpha}}H_{ji}(k_1)\Bigr|_{k_1=x_1p}
\Biggr]k_{1\perp}^{\alpha}\Biggr\}.
\label{eq-H1_sum}
\eea
We iteratively used the third relation in Eq. (\ref{WTI-nonpole2}) for the first equation.
Substituting Eq. (\ref{eq-H1_sum}) into Eq. (\ref{eq-W1_2}), we obtain the final result 
\beq
W^{(1)}
&=&p^+\int{dx}\int{dy_1\over 2\pi}\,e^{ixp^+y_1^-}
\la pS_{\perp}|\bar{\psi}_j(0)\Bigl[ig\int^0_{y_1^-}{dy_2^-}\,A^{n}(y_2^-)\Bigr]
\psi_i(y_1^-)|pS_{\perp}\ra H_{ji}(xp)
\nonumber\\
&&-ip^+\int{dx_1}\int{dx_2}\int{dy_1^-\over 2\pi}\int{dy_2^-\over 2\pi}
\,e^{ix_1p^+y_1^-}e^{i(x_2-x_1)p^+y_2^-}
\la pS_{\perp}|\bar{\psi}_j(0) gF^{\alpha n}(y_2^-)
\psi_i(y_1^-)|pS_{\perp}\ra
{H_{ji,\alpha}(x_1p,x_2p) \over x_2-x_1-i\epsilon}
\nonumber\\
&&+ip^+\int{dx}\int{dy_1^-\over 2\pi}\,e^{i xp^+y_1^-}
\la pS_{\perp}|\bar{\psi}_j(0)\Bigl[ig\int^0_{y_1^-}{dy_2^-}\,A^{n}(y_2^-)\Bigr]
\partial_{\perp}^{\alpha}\psi_i(y_1^-)|pS_{\perp}\ra
{\partial\over \partial k^{\alpha}}H_{ji}(k)\Bigr|_{k=xp}
\nonumber\\
&&+ip^+\int{dx}\int{dy_1^-\over 2\pi}\,e^{ixp^+y_1^-}
\la pS_{\perp}|\bar{\psi}_j(0)ig\int^{\infty}_{y_1^-}{dy_2^-}
F^{\alpha n}(y_2^-) \psi_i(y_1^-)|pS_{\perp}\ra
{\partial\over \partial k^{\alpha}}H_{ji}(k)\Bigr|_{k=xp}
\nonumber\\
&&+ip^+\int{dx}\int{dy_1^-\over 2\pi}
\,e^{ixp^+y_1^-}\la pS_{\perp}|\bar{\psi}_j(0)\Bigl[-igA_{\perp}^{\alpha}(y_1^-)\Bigr]
\psi_i(y_1^-)|pS_{\perp}\ra{\partial \over \partial k^{\alpha}}H_{ji}(k)\Bigr|_{k=xp}.
\label{eq-W1_fin}
\eeq
Summing over all twist-3 contributions in the diagrams in Figs. \ref{fig1} and \ref{fig3}, represented by Eqs. (\ref{eq-W0}, \ref{eq-W1_fin}), respectively, we can construct the gauge-invariant expression
\beq
W=\int dx\,{\rm Tr}[M(x)H(xp)]+i\int dx
\,{\rm Tr}\Biggl[M^{\alpha}_{\partial}(x){\partial\over \partial k^{\alpha}}H(k)\Bigr|_{k=xp}\Biggr]
\nonumber\\
-i\int dx_1\int dx_2{1\over x_2-x_1-i\epsilon}
{\rm Tr}[M_F^{\alpha}(x_1,x_2)H_{\alpha}(x_1p,x_2p)],
\label{nonpolew}
\eeq
where the matrices are given by
\bea
M_{ij}(x)=&p^+\int{dy_1^-\over 2\pi}e^{ixp^+y_1^-}\la pS_{\perp}|\bar{\psi}_j(0)\psi_i(y_1^-)|pS_{\perp}\ra,
\\
M^{\alpha}_{ij,\partial}(x)=&p^+\int{dy_1^-\over 2\pi}e^{ixp^+y_1^-}
\la pS_{\perp}|\bar{\psi}_j(0)D_{\perp}^{\alpha}(y_1^-)\psi_i(y_1^-)|pS_{\perp}\ra
\nonumber\\
&+p^+\int{dy_1^-\over 2\pi}e^{ixp^+y_1^-}\,
\la pS_{\perp}|\bar{\psi}_j(0)ig\Bigl[\int^{\infty}_{y_1^-}dy_2^- F^{\alpha n}(y_2^-)\Bigr]
\psi_i(y_1^-)|pS_{\perp}\ra,
\nonumber\\
=&-i{M_N\over 2}\epsilon^{\alpha \bar n nS_{\perp}}(\slash{p})_{ij}f_{1T}^{\perp(1)}(x)+\cdots,
\eea
where operator definition of $f_{1T}^{\perp(1)}(x)$ is
\bea
f_{1T}^{\perp(1)}(x) =\Bigl({-i\over 2M_N}\Bigr)
\int{dy_1^-\over 2\pi}e^{ixp^+y_1^-}\la pS_{\perp}|\bar{\psi}(0)\slash{n}
\epsilon^{\alpha \bar n nS_{\perp}}\Bigl(D_{\perp\alpha}(y_1^-)
+ig\Bigl[\int^{\infty}_{y_1^-}dy_2^- F_\alpha^{\ n}(y_2^-)\Bigr]\Bigr)
\psi(y_1^-)|pS_{\perp}\ra.
\eea
The definition of $M_F^{\alpha}(x_1,x_2)$ and its decomposition are already introduced in Eq. (\ref{eq-MF_def}).
In the present case, the first term in Eq. (\ref{nonpolew}) can't give a twist-3 contribution because the spin projection
$\gamma_{\alpha}\epsilon^{\alpha \bar n nS_{\perp}}$ is forbidden by $PT$-invariance.
Therefore, we can eliminate the first term in Eq. (\ref{nonpolew}) and rewrite the twist-3 hadronic part 
as
\beq
W&=&{M_N\over 2}\epsilon^{\alpha \bar n nS_{\perp}}
\Biggl\{\int dx\,f_{1T}^{\perp(1)}(x) {\rm Tr}\Bigl[\slash{p}{\partial\over \partial k^{\alpha}}H(k)\Bigr|_{k=xp}\Bigr]
\nonumber\\
&&+i\int dx_1\int dx_2\,T_{q,F}(x_1,x_2){1\over x_2-x_1-i\epsilon}
{\rm Tr}[\slash{p}H_{\alpha}(x_1p,x_2p)]\Biggr\}.
\label{nonpolew_2}
\eeq
In the new method presented above, we only needed the well-defined relations Eq. (\ref{WTI-nonpole2}) to construct the gauge-invariant matrix elements. We find that the difficulty associated with the relation Eq. (\ref{relationSGP}) in the conventional calculation was removed. This is one of the advantages in the new method. Another advantage is that, by using Eq. (\ref{nonpolew_2}) and the discussion in Appendix. B, we don't need to calculate the derivative of the hard part over the momentum $k(k_i)$, this will significantly reduce the complexity of twist-3 calculation, in particularly for higher order calculations. 

%%%%%%%%%%%%%%%%%%%%%%
\subsection{SIDIS at $\mathcal O(\alpha_S)$}
In this subsection, we show in detail the calculation of hadronic part for SIDIS at $\mathcal{O}(\alpha_s)$.
We factor out the on-shell $\delta$-function from the hard partonic part,
\beq
H(k)=&&\bar{H}(k)(2\pi)\delta\left[(k+q-p_c)^2\right],
\\
H_{\alpha}(x_1p,x_2p)=&&\bar{H}^L_{\alpha}(x_1p,x_2p)(2\pi)\delta\left[(x_2p+q-p_c)^2\right]
+\bar{H}^R_{\alpha}(x_1p,x_2p)(2\pi)\delta\left[(x_1p+q-p_c)^2\right],
\eeq
where $\bar{H}^L_{\alpha}(x_1p,x_2p)$ is given by a sum of 12 diagrams in Fig. \ref{fig4} and 
$\bar{H}^R_{\alpha}(x_1p,x_2p)$ is its complex conjugate. The derivative of $H(k)$ over $k$ can be converted to that over the standard Mandelstam variables $\hat s, \hat t, \hat u$. For details, see
Appendix B.  Then we can calculate Eq. (\ref{nonpolew_2}) as
\beq
W&=&\pi M_N\int {dx\over x}\delta\Bigl[(xp+q-p_c)^2\Bigr]
\Biggl\{x{df_{1T}^{\perp(1)}(x) \over dx}\left(\epsilon^{q\bar{n}nS_{\perp}}-\epsilon^{p_c\bar{n}nS_{\perp}}
\right){2\over \hat{u}}\hat{\sigma}(\hat s, \hat t, \hat u)
+f_{1T}^{\perp(1)}(x)\Biggl[\Bigl((\hat{s}+Q^2)\epsilon^{p_c\bar{n}nS_{\perp}}
+\hat{t}\epsilon^{q\bar{n}nS_{\perp}}\Bigr)
\nonumber\\
&&\times{2\over \hat{u}}
\left({\partial\over \partial\hat{t}}-{\partial\over \partial\hat{s}}\right)
\hat{\sigma}(\hat s, \hat t, \hat u)
-\Bigl(\epsilon^{q\bar{n}nS_{\perp}}-\epsilon^{p_c\bar{n}nS_{\perp}}\Bigr){2\over \hat{u}}
\hat{\sigma}(\hat s, \hat t, \hat u)
-\epsilon^{\alpha \bar{n}nS_{\perp}}{\rm Tr}[\gamma_{\alpha}\bar{H}(xp)]\Biggr]
\nonumber\\
&&+i\epsilon^{\alpha\bar{n}nS_{\perp}}\int dx'\,T_{q,F}(x',x)\left[{1\over x-x'-i\epsilon}
{\rm Tr}[x\slash{p}\bar{H}^L_{\alpha}(x'p,xp)]-{1\over x-x'+i\epsilon}{\rm Tr}[x\slash{p}\bar{H}^R_{\alpha}(xp,x'p)]\right]\Biggr\},
\label{nonpole}
\eeq
where $\hat{\sigma}(\hat s, \hat t, \hat u)$ is the $2\to 2$ partonic cross section in SIDIS. For $q\gamma^*\to qg$ channel, it reads
\beq
\hat{\sigma}(\hat s, \hat t, \hat u)\equiv {\rm Tr}[x\slash{p}\bar{H}(xp)]
=-8C_FQ^2{(\hat{s}+\hat{t})^2+(\hat{t}+\hat{u})^2\over \hat{s}\hat{u}}.
\label{twist-2}
\eeq
Notice that, for convenience, we have changed the notation $x_1\to x',~ x_2\to x$ in $\bar{H}^L(x'p,xp)$, and $x_1\to x,~ x_2\to x'$ in $\bar{H}^R(x'p,xp)$ in order to factor out the common delta function
$\delta\Bigl[(xp+q-p_c)^2\Bigr]$. We discuss the gauge- and Lorentz-invariances of the hard cross sections associated with 
$f_{1T}^{\perp (1)}(x)$. The hard cross section with the nonderivative function $f_{1T}^{\perp (1)}(x)$
is not apparently gauge-invariant because of the term $\epsilon^{\alpha pnS_{\perp}}{\rm Tr}[\gamma_{\alpha}\bar{H}(xp)]$.
The gauge-invariance requires the unpolarized spin projection $x\slash{p}$ with 
$\bar{H}(xp)$ like Eq. (\ref{twist-2}). On the other hand, the hard cross section associated with the derivative function ${d\over dx}f_{1T}^{\perp (1)}(x)$ is not Lorentz-invariant. The vector $n$ in the parametrization 
(\ref{eq-MF_def}) satisfies $\bar{n}\cdot n=1$ and $n^2=0$. These conditions are not enough 
to uniquely determine the form of $n$ and there are two possible choices in SIDIS,
\beq
{n^{\alpha}}={p^+\over p\cdot p_c}p_c^{\alpha}
\hspace{5mm}{\rm or}\hspace{5mm}
{n^{\alpha}}={p^+\over p\cdot p_c}p_c^{\alpha}+{2 p^+ p_c\cdot q\over 2(p_c\cdot q)(p\cdot q)
+Q^2(p\cdot p_c)}\left(q^{\alpha}-{p\cdot q\over p\cdot p_c}p_c^{\alpha}\right).
\eeq
We can check that the coefficient $(\epsilon^{q\bar n nS_{\perp}}-\epsilon^{p_c\bar n nS_{\perp}})$
of ${d\over dx}f_{1T}^{\perp (1)}(x)$ depends on the choice of $n$. This ambiguity of the cross section
is physically interpreted as the frame-dependence because the spatial components of $n$ is determined
so that it has the opposite direction of the momentum $p$ as $\vec{n}=-\vec{p}/p^+$. From the 
requirement of the frame-independence, the cross section has to be proportional to the factor
$\left[(\hat{s}+Q^2)\epsilon^{p_c\bar n nS_{\perp}}+\hat{t}\epsilon^{q\bar n nS_{\perp}}\right]$ as already shown in the cross section (\ref{result_pole}) derived by the conventional pole method.
We will show later that the gauge- and Lorenz-invariances of the cross section are guaranteed
by using the relations
\beq
f_{1T}^{\perp(1)}(x)&=&\pi T_{q,F}(x,x),
\label{GIR}
\\
{d\over dx}f_{1T}^{\perp(1)}(x)&=&\pi {d\over dx}T_{q,F}(x,x),
\label{LIR}
\eeq
which enable us to express the cross section only in terms of $T_{q,F}(x',x)$ as in the case 
of the conventional calculation. One can find the derivation of these relations in Appendix A.
%%%%%%%%%%%%%%%%%%%%%%%%%%%%%%%%%%%%%%%%%%%%%%%%%%%%%%%%%%%%%%%%%%%%%%%%%%%%%%%
\begin{figure}[h]
\begin{center}
  \includegraphics[height=4cm,width=12cm]{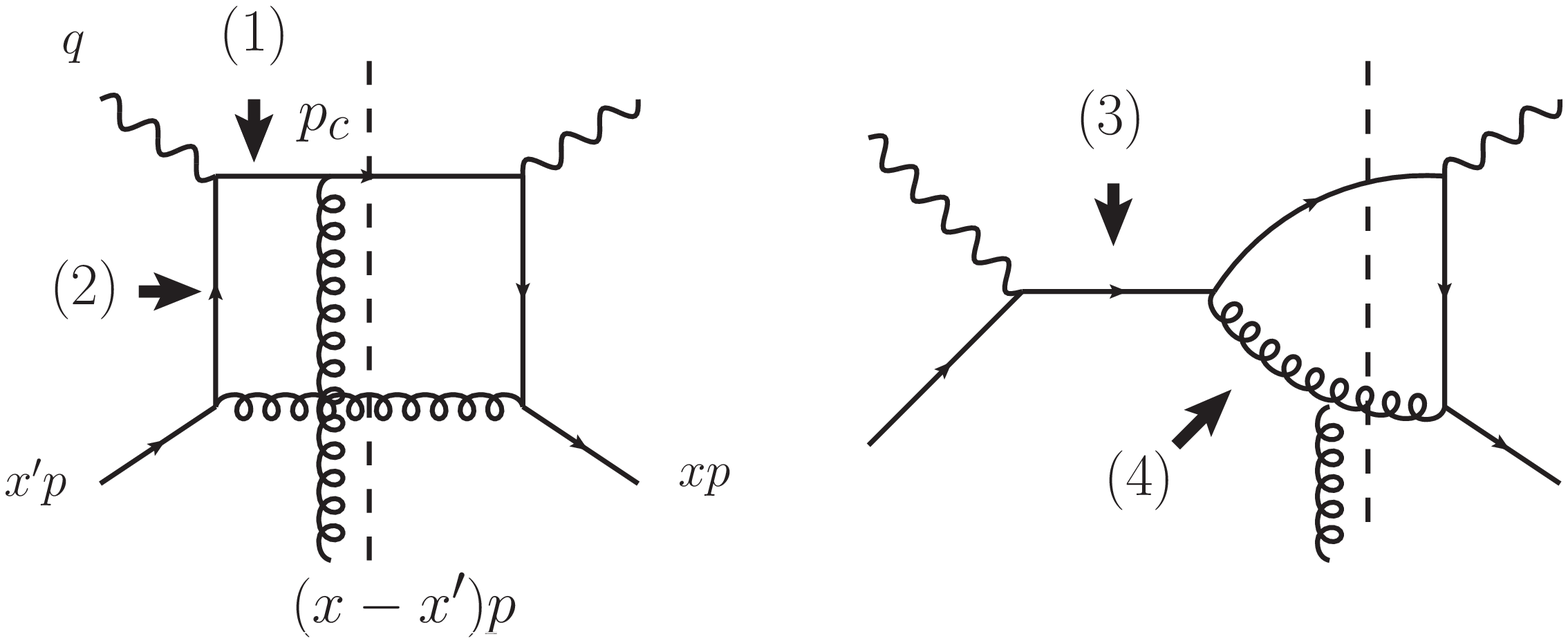}\hspace{1cm}
\end{center}
 \caption{Typical diagrams including $x'$-dependent propagators. 
Calculation for the propagators (1)-(4) are shown in Eq. (\ref{x'dep-prop}).}
\label{fig5}
\end{figure}
%%%%%%%%%%%%%%%%%%%%%%%%%%%%%%%%%%%%%%%%%%%%%%%%%%%%%%%%%%%%%%%%%%%%%%%%%%%%%%% 

Now we show how to calculate the hard partonic part $\bar{H}^L(x'p,xp)$. There are four types of $x'$-dependences in Feynman gauge. Fig. 5 shows typical diagrams including $x'$-dependent propagators.
Each propagator can be calculated as follows: 
\beq
&&{\rm propagator\ (1):}\hspace{5mm}{\slash{p}_c-(x-x')\slash{p}\over [p_c-(x-x')p]^2+i\epsilon}=-{1\over \hat{t}}x\slash{p}+{x\over x-x'-i\epsilon}{1\over\hat{t}}\slash{p}_c,
\nonumber\\
&&{\rm propagator\ (2):}\hspace{5mm}{\slash{p}_c-(x-x')\slash{p}-\slash{q}\over [p_c-(x-x')p-q]^2+i\epsilon}={1\over \hat{u}}x\slash{p}-{x\over x'-i\epsilon}{1\over\hat{u}}(x\slash{p}+\slash{q}-\slash{p}_c),
\nonumber\\
&&{\rm propagator\ (3):}\hspace{5mm}{x'\slash{p}+\slash{q}\over [x'p+q]^2+i\epsilon}={1\over \hat{s}+Q^2}x\slash{p}+{x\over x'-x_B+i\epsilon}{1\over \hat{s}+Q^2}
[x_B\slash{p}+\slash{q}],
\nonumber\\
&&{\rm propagator\ (4):}
\nonumber\\
&&{V_{\alpha\rho\tau}\Bigl((x-x')p,-xp-q+p_c,x'p+q-p_c\Bigr)\over [x'p+q-p_c]^2+i\epsilon}
={1\over \hat{u}}
\Bigl(xp_{\tau}g_{\alpha\rho}+xp_{\alpha}g_{\rho\tau}
-2xp_{\rho}g_{\alpha\tau}\Bigr)
\nonumber\\
&&+{x\over x-x'-i\epsilon}{1\over \hat{u}}
\Bigl[(xp+q-p_c)_{\tau}g_{\alpha\rho}-2(xp+q-p_c)_{\alpha}g_{\rho\tau}
+(xp+q-p_c)_{\rho}g_{\alpha\tau}\Bigr],
\label{x'dep-prop}
\eeq
where $V_{\alpha\rho\tau}$ comes from the 3-gluon vertex. We can find that all $x'$-dependence
appear only in the denominators, $x-x'-i\epsilon$, $x'-i\epsilon$ and $x'-x_B+i\epsilon$. 
Products of two denominators can be disentangled as
\beq
{x\over x-x'-i\epsilon}{x\over x'-i\epsilon}&=&{x\over x-x'-i\epsilon}+{x\over x'-i\epsilon},
\nonumber\\
{x\over x-x'-i\epsilon}{x\over x'-x_B+i\epsilon}&=&{\hat{s}+Q^2\over \hat{s}}
\Bigr({x\over x-x'-i\epsilon}+{x\over x'-x_B+i\epsilon}\Bigr).
\eeq
From the above discussion, we can conclude that the part of the cross section with $\bar{H}^L_{\alpha}(x'p,xp)$ is given by
\beq
&&\epsilon^{\alpha\bar{n}nS_{\perp}}\int dx'T_{q,F}(x',x){1\over x-x'-i\epsilon}
{\rm Tr}[x\slash{p}\bar{H}^L_{\alpha}(x'p,xp)]
\nonumber\\
&=&\int dx'T_{q,F}(x',x)\Bigl[
{1\over x-x'-i\epsilon}H_{F1}
+{x\over (x-x'-i\epsilon)^2}H_{F2}
+{1\over x'-i\epsilon}H_{F3}
+{1\over x'-x_B+i\epsilon}H_{F4}
\Bigr].
\eeq
All the hard parts $H_{Fi}$ are independent of $x'$.
We can repeat the same discussion on $\bar{H}^R_{\alpha}(xp,x'p)$. Then we can calculate each hard partonic cross section and obtain the following result for the hadronic part.
\beq
W
&=&\pi M_N\int {dx\over x}\delta\Bigl[(xp+q-p_c)^2\Bigr]
\Biggl\{x{df_{1T}^{\perp(1)}(x) \over dx}\Bigl(\epsilon^{q\bar{n}nS_{\perp}}-\epsilon^{p_c\bar{n}nS_{\perp}}
\Bigr){2\over \hat{u}}\hat{\sigma}(\hat s, \hat t, \hat u)
\nonumber\\
&&+\Bigl[(\hat{s}+Q^2)\epsilon^{p_c\bar{n}nS_{\perp}}
+\hat{t}\epsilon^{q\bar{n}nS_{\perp}}\Bigr]f_{1T}^{\perp(1)}(x)\hat{\sigma}_{ND'}
\nonumber\\
&&+i\int dx'T_{q,F}(x',x)\biggl[
\left({1\over x-x'-i\epsilon}-{1\over x-x'+i\epsilon}\right)H_{F1}
+\left({x\over (x-x'-i\epsilon)^2}-{x\over (x-x'+i\epsilon)^2}\right)H_{F2}
\nonumber\\
&&+\Bigl({1\over x'-i\epsilon}-{1\over x'+i\epsilon}\Bigr)H_{F3}
+\Bigl({1\over x'-x_B+i\epsilon}-{1\over x'-x_B-i\epsilon}\Bigr)H_{F4}
\biggr]
\Biggr\},
\label{nonpole_result}
\eeq
where the hard cross sections are given by
\beq
\hat{\sigma}_{ND'}&=&16C_FQ^2{Q^2\hat{t}-\hat{t}^2-\hat{t}\hat{u}-\hat{u}^2\over \hat{s}^2\hat{u}^2},
\nonumber\\
H_{F1}&=&\Bigl[(\hat{s}+Q^2)\epsilon^{p_c\bar{n}nS_{\perp}}
+\hat{t}\epsilon^{q\bar{n}nS_{\perp}}\Bigr]\left(-{1\over 2}\hat{\sigma}_{ND}+{1\over 2}\hat{\sigma}_{ND'}\right),
\nonumber\\
H_{F2}&=&\Bigl[(\hat{s}+Q^2)\epsilon^{p_c\bar{n}nS_{\perp}}
+\hat{t}\epsilon^{q\bar{n}nS_{\perp}}\Bigr]\hat{\sigma}_{D}-\Bigl(\epsilon^{q\bar{n}nS_{\perp}}-\epsilon^{p_c\bar{n}nS_{\perp}}
\Bigr){2\over \hat{u}}\hat{\sigma}(\hat s, \hat t, \hat u),
\nonumber\\
H_{F3}&=&-{1\over 2}\Bigl[(\hat{s}+Q^2)\epsilon^{p_c\bar{n}nS_{\perp}}+\hat{t}\epsilon^{q\bar{n}nS_{\perp}}\Bigr]
\hat{\sigma}_{SFP},
\nonumber\\
H_{F4}&=&{1\over 2}\Bigl[(\hat{s}+Q^2)\epsilon^{p_c\bar{n}nS_{\perp}}+\hat{t}\epsilon^{q\bar{n}nS_{\perp}}\Bigr]
\hat{\sigma}_{HP}.
\eeq
$\hat\sigma_{ND}, \hat\sigma_D, \hat\sigma_{SFP},~\hat\sigma_{HP}$ can be found in Eq. (\ref{eq-sigmahat}). Since $H_{Fi}$ are all independent of $x'$, the $x'$-integration only involves $T_{q,F}(x',x)$ and the propagators. 
Then we can perform $x'$-integration in Eq. (\ref{nonpole_result}) as
\beq
\int dx'\Bigl({1\over x-x'-i\epsilon}-{1\over x-x'+i\epsilon}\Bigr)T_{q,F}(x',x)
&=&2\pi iT_{q,F}(x,x),
\nonumber\\
\int dx'\Bigl[{x\over (x-x'-i\epsilon)^2}-{x\over (x-x'+i\epsilon)^2}\Bigr]T_{q,F}(x',x)
&=&-\pi ix{d\over dx}T_{q,F}(x,x),
\nonumber\\
\int dx'\Bigl({1\over x'-i\epsilon}-{1\over x'+i\epsilon}\Bigr)T_{q,F}(x',x)
&=&2\pi iT_{q,F}(0,x),
\nonumber\\
\int dx'\Bigl({1\over x'-x_B+i\epsilon}-{1\over x'-x_B-i\epsilon}\Bigr)T_{q,F}(x',x)
&=&-2\pi iT_{q,F}(x_B,x),
\eeq
where we have used the symmetric property of Qiu-Sterman function $T_{q,F}(x',x)=T_{q,F}(x,x')$ in the integration of the double pole coefficient. Substituting these relations into Eq. (\ref{nonpole_result}) and using Eqs. (\ref{GIR},\ref{LIR}),
we can finally derive the transverse polarized cross section in SIDIS based on the new method as
\beq
&&{d^4\Delta\sigma\over dx_Bdydz_hdP_{h\perp}}
\nonumber\\
&=&{\pi M_N\alpha^2_{em}\alpha_s\over 8z_hx_B^2S^2_{ep}Q^2}\sum_qe^2_q\int{{dz\over z^2}}D_{q\to h}(z)\int{dx\over x}
\delta\Bigl[(xp+q-p_c)^2\Bigr]\Bigl((\hat{s}+Q^2)\epsilon^{p_c\bar{n}nS_{\perp}}+\hat{t}\epsilon^{q\bar{n}nS_{\perp}}\Bigr)
\nonumber\\
&&\times\Bigl[x{d\over dx}T_{q,F}(x,x)\hat{\sigma}_{D}+T_{q,F}(x,x)\hat{\sigma}_{ND}
+T_{q,F}(0,x)\hat{\sigma}_{SFP}
+T_{q,F}(x_B,x)\hat{\sigma}_{HP}
\Bigr].
\eeq
This is exactly the same with the result of the conventional calculation (\ref{result_pole}).
We would like to emphasize that the cross section is never gauge- and Lorentz-invariant 
if the kinematical function $f_{1T}^{\perp(1)}(x)$ and Qiu-Sterman function $T_{q,F}(x,x)$ are
independent of each other. The relation between them is needed for the physically acceptable result.  

In the end, we make a comment on the generality of our result. 
We only  considered the metric part $L^{\mu\nu}\simeq -Q^2g^{\mu\nu}$  
in our calculation so that one can easily follow the calculation
and clearly see the difference between two calculation methods. It's a natural question whether 
the consistency holds when we consider the full leptonic tensor shown in Eq. (\ref{leptonic}). The conventional way to calculate the cross section in SIDIS is that we expand the hadronic tensor in terms of orthogonal bases.
The symmetric part of the tensor $W^{\mu\nu}$ has 10 independent components and one of them is fixed by the condition $q_{\mu}W^{\mu\nu}=0$. Then $W^{\mu\nu}$ can be expanded by 9 independent bases as
\beq
W^{\mu\nu}=\sum_{i=1}^{9}(W^{\rho\sigma}\tilde{{\cal V}}_{i\,\rho\sigma}){\cal V}^{\mu\nu}_i.
\eeq
One can find the explicit forms of ${\cal V}_{i\,\rho\sigma}$ and $\tilde{{\cal V}}^{\mu\nu}_i$ in Ref.~\cite{Meng:1991da}. 
Then the contracted form with $L^{\mu\nu}$ is rewritten as
\beq
L^{\mu\nu}W_{\mu\nu}=
\sum_{i=1}^{9}(L^{\mu\nu}{\cal V}_{i\,\mu\nu})(W^{\rho\sigma}\tilde{{\cal V}}_{i\,\rho\sigma}).
\eeq
This equation means that the calculation with the full leptonic tensor $L^{\mu\nu}$ results in the calculation of the hard cross sections $W^{\rho\sigma}\tilde{{\cal V}}_{i\,\rho\sigma}$.
Three tensors $\tilde{{\cal V}}^{\mu\nu}_{5,6,7}$ are irrelevant to our study because they are pure imaginary.
We verified that the consistency between the two methods holds for all 6 hard cross sections
($i=1,2,3,4,8,9$). This result shows that the consistency holds for the full leptonic tensor and  enhances the generality of our result.

%----------------- Summary ------------------------

\section{Summary}

We proposed the new nonpole calculation method for the Sivers effect in the twist-3 cross section and confirmed the consistency with the conventional pole calculation. We found out that the relation $f_{1T}^{\perp(1)}(x)=\pi T_{q,F}(x,x)$ is very important to guarantee the gauge- and Lorentz-invariances of the final result. We reproduced this relation without introducing the definition of the TMD Sivers function. The importance of Eq. (\ref{GIR}) has been mainly discussed in the context of the matching
between the TMD factorization and the collinear twist-3 factorization frameworks\cite{Scimemi:2018mmi,Scimemi:2019gge}. Our calculation showed that this was also important for the gauge- and Lorentz-invariances of the twist-3 physical observables for the Sivers effect. This result provides a new perspective on the relation. The same technique can be also applied to the gluon Sivers function and the twist-3 gluon distribution functions~\cite{Koike:2011mb}. The relation between them is relatively nontrivial compared to the quark functions. From the requirement of the gauge- and Lorentz-invariances of the twist-3 cross section, we can derive a similar relation with Eq. (\ref{GIR}) for the gluon distribution functions.

One of the advantages in the new non-pole calculation method is that we don't need to prove Eq. (\ref{relationSGP}) for the SGP contribution as required in conventional pole method, which can be only checked through diagram by diagram calculation.
It's known that this relation may not be hold when the description of the fragmentation part is changed to other framework such as NRQCD for heavy quarkonium production. In the new method, we never separate the pole contributions and then no singularity arises from the relation associated with WTI. Our new method will extend the applicability of the collinear twist-3 framework.

In the new method, one does not need to perform derivatives over the initial parton's transverse momentum in the calculation of Feynman diagrams. We can anticipate that a lot of propagators depend on  the initial parton's momentum in higher order diagrams. The direct operation of 
the derivatives is highly complicated task. Our method could significantly reduce this complexity as discussed in Sec. III. As mentioned just below Eq.~(\ref{WTI-nonpole2}), the WTI doesn't
change for the higher order diagrams as long as the gauge invariance is preserved. Most of our results are available without change for the higher 
order cross section in SIDIS. A set of equations derived in this paper could be useful 
to derive the first next-leading order cross section for the SSA in $ep$-collision 
which could be measured at Electron-Ion-Collider in the near future.

We expect the new method presented in this manuscript can be extended to higher-twist calculation, which becomes one of the standard method to investigate the nontrivial nuclear effect in heavy ion collisions \cite{Kang:2008us,Kang:2011bp,Xing:2012ii,Kang:2013ufa,Kang:2014hha}. As we don't need to perform derivatives over the initial parton's transverse momentum in the new non-pole method, we expect the new approach will be of great use in performing next-to-leading order calculation at higher twist, in which the conventional collinear expansion caused ambiguity in setting up the initial parton's kinematics \cite{Kang:2014ela, Kang:2016ron}, this ambiguity can be resolved in the new non-pole method.

\

%%%%%%%%%%%%%%%%%%%%%%%%%%%%%%%%%%%%%%%%%%%%%%%%%%%%

\appendix

\section{Twist-3 quark-gluon correlation functions}

\noindent
\subsection{Definition of the twist-3 functions}

We introduce the definition of all relevant twist-3 functions for the transversely polarized proton~\cite{Kanazawa:2015ajw,Eguchi:2006qz}.

\

\noindent
\underline{D-type dynamical function}

\beq
M^{\alpha}_{ij,D}(x_1,x_2)&=&(p^+)^2\int{dy_1^-\over 2\pi}\int{dy_2^-\over 2\pi}e^{ix_1p^+y_1^-}e^{i(x_2-x_1)p^+y_2^-}
\la pS_{\perp}|\bar{\psi}_j(0)[0,y_2^-]D_{\perp}^{\alpha}(y_2^-)[y_2^-,y_1^-]\psi_i(y_1^-)|pS_{\perp}\ra
\nonumber\\
&=&-{M_N\over 2}\epsilon^{\alpha \bar{n}nS_{\perp}}(\slash{p})_{ij}T_{q,D}(x_1,x_2)+\cdots,
\eeq
where $D_{\perp}^{\alpha}(y_2^-)=\partial_{\perp}^{\alpha}-igA_{\perp}^{\alpha}(y_2^-)$,
$[0,y_2^-]$ is the Wilson line
\beq
[0,y_2^-]=P{\rm exp}\Bigl(ig\int^{0}_{y_2^-}dy^-\,A^n(y^-)\Bigr).
\eeq
The D-type function $T_{q,D}(x_1,x_2)$ is real and antisymmetric $T_{q,D}(x_1,x_2)=-T_{q,D}(x_2,x_1)$.

\

\noindent
\underline{Kinematical function}

\beq
M^{\alpha}_{ij,\partial}(x)&=&p^+\int{dy_1^-\over 2\pi}e^{ixp^+y_1^-}
\la pS_{\perp}|\bar{\psi}_j(0)[0,y^-_1]D_{\perp}^{\alpha}(y_1^-)\psi_i(y_1^-)|pS_{\perp}\ra
\nonumber\\
&&+p^+\int{dy_1^-\over 2\pi}e^{ixp^+y_1^-}\,
\la pS_{\perp}|\bar{\psi}_j(0)ig\Bigl[\int^{\infty}_{y_1^-}dy_2^- [0,y^-_2]
F^{\alpha n}(y_2^-)[y^-_2,y^-_1]\Bigr]
\psi_i(y_1^-)|pS_{\perp}\ra,
\nonumber\\
&=&-i{M_N\over 2}\epsilon^{\alpha \bar n nS_{\perp}}(\slash{p})_{ij}f_{1T}^{\perp(1)}(x)+\cdots.
\label{kinematical}
\eeq 
By using the translation invariance~\cite{Kanazawa:2015ajw},
\beq
&&\la pS_{\perp}|\bar{\psi}_j(0)\overleftarrow{D}_{\perp}^{\alpha}(0)[0,y_1^-]\psi_i(y_1^-)|pS_{\perp}\ra
+\la pS_{\perp}|\bar{\psi}_j(0)[0,y_1^-]D_{\perp}^{\alpha}(y_1^-)\psi_i(y_1^-)|pS_{\perp}\ra
\nonumber\\
&&+\int^{0}_{y_1^-}{dy_2^-\over 2\pi}\la pS_{\perp}|\bar{\psi}_j(0)[0,y_2^-]igF^{\alpha n}(y_2^-)
[y_2^-,y_1^-]\psi_i(y_1^-)|pS_{\perp}\ra=0,
\eeq
we can show $M_{\partial}^*(x)=-M_{\partial}(x)$ and therefore
$f_{1T}^{\perp(1)}(x)$ is real function. The kinematical function $f_{1T}^{\perp(1)}(x)$ has another 
definition using the quark TMD correlator. Here we recall the definition of the quark Sivers function
~\cite{Bacchetta:2006tn},
\beq
M_{ij}(x,p_T)&=&\int{dy^-\over 2\pi}\int {d^2\xi_T\over 2\pi}e^{ixp^+y^-}e^{ip_T\cdot \xi_T}
\la pS_{\perp}|\bar{\psi}_j(0)[0,\infty^-][\infty^-,\infty^-+\xi_T]
\nonumber\\
&&\times[\infty^-+\xi_T,y^-+\xi_T]
\psi_i(y^-+\xi_T)|pS_{\perp}\ra
\nonumber\\
&=&-{1\over 2M_N}f_{1T}^{\perp}(x,p_T)\epsilon^{p_T\bar{n}\mu S_{\perp}}\gamma_{\mu}+\cdots.
\eeq
We can find a relation between the first moment of $M(x,p_T)$ and the correlator of the kinematical 
function $M^{\alpha}_{\partial}(x)$,
\beq
&&\int d^2p_T\,p^{\alpha}_TM_{ij}(x,p_T)
\nonumber\\
&=&\int{dy^-\over 2\pi}\int {d^2\xi_T\over 2\pi}e^{ixp^+y^-}
\Bigl(-i{\partial\over \partial \xi_{T\alpha}}\Bigr)
e^{ip_T\cdot \xi_T}
\la pS_{\perp}|\bar{\psi}_j(0)[0,\infty^-][\infty^-,\infty^-+\xi_T][\infty^-+\xi_T,y^-+\xi_T]
\psi_i(y^-+\xi_T)|pS_{\perp}\ra
\nonumber\\
&=&i\int{dy^-\over 2\pi}e^{ixp^+y^-}
\la pS_{\perp}|\bar{\psi}_j(0)D_{\perp}^{\alpha}(y^-)\psi_i(y^-)|pS_{\perp}\ra
+i\int{dy^-\over 2\pi}e^{ixp^+y^-}\,
\la pS_{\perp}|\bar{\psi}_j(0)ig\Bigl[\int^{\infty}_{y^-}dy_2^- F^{\alpha n}(y_2^-)\Bigr]
\psi_i(y^-)|pS_{\perp}\ra
\nonumber\\
&=&{i\over p^+}M^{\alpha}_{ij,\partial}(x).
\eeq
Then $f_{1T}^{\perp(1)}(x)$ can be expressed by the first moment of the quark Sivers function~\cite{Boer:2003cm,Kang:2011hk}.
\beq
f_{1T}^{\perp(1)}(x)=\int d^2p_T\,{|p_{T}|^2\over 2M_N^2}f_{1T}^{\perp }(x,p_T).
\eeq
The matching between TMD functions and collinear functions itself is an active research subject in perturbative QCD
phenomenology. One can find recent developments in Refs.~\cite{Scimemi:2018mmi,Scimemi:2019gge} and references therein.

\

\noindent
\underline{F-type dynamical function}

\beq
M_{ij, F}^{\alpha}(x_1,x_2)&=&p^+ \int{dy_1^-\over 2\pi}\int{dy_2^-\over 2\pi}
\,e^{i x_1p^+y_1^-}e^{i(x_2-x_1)p^+y_2^-}
\la pS_{\perp}|\bar{\psi}_j(0)[0,y^-_2]gF^{\alpha n}(y_2^-)[y^-_2,y^-_1]\psi_i(y_1^-)|pS_{\perp}\ra 
\nonumber\\
&=&-{M_N\over 2}\epsilon^{\alpha \bar n nS_{\perp}}(\slash{p})_{ij}T_{q,F}(x_1,x_2)+\cdots,
\label{Ftype}
\eeq
where the F-type function $T_{q,F}(x_1,x_2)$ is real and symmetric $T_{q,F}(x_1,x_2)=T_{q,F}(x_2,x_1)$.

\subsection{Relation among the functions}

We can derive an operator identity among the three types of correlators~\cite{Eguchi:2006qz}.
In order to derive the relation, we use the identity for the 
$D_{\perp}^{\alpha}(y_2^-)[y_2^-,y_1^-]$ in $M^{\alpha}_D(x_1,x_2)$,
\beq
&&D_{\perp}^{\alpha}(y_2^-)[y_2^-,y_1^-]
\nonumber\\
&=&[y_2^-,y_1^-]D_{\perp}^{\alpha}(y_1^-)
+i\int^{y_2^-}_{y_1^-}dy_3^-[y_2^-,y_3^-]gF^{\alpha n}(y_3^-)[y_3^-,y_1^-]
\nonumber\\
&=&[y_2^-,y_1^-]D_{\perp}^{\alpha}(y_1^-)
+i\int^{\infty}_{y_1^-}dy_3^-[y_2^-,y_3^-]gF^{\alpha n}(y_3^-)[y_3^-,y_1^-]
-i\int^{\infty}_{y_2^-}dy_3^-[y_2^-,y_3^-]gF^{\alpha n}(y_3^-)[y_3^-,y_1^-]
\nonumber\\
&=&[y_2^-,y_1^-]D_{\perp}^{\alpha}(y_1^-)
+i\int^{\infty}_{y_1^-}dy_3^-[y_2^-,y_3^-]gF^{\alpha n}(y_3^-)[y_3^-,y_1^-]
-i\int^{\infty}_{-\infty}dy_3^-\theta(y_3^--y_2^-)[y_2^-,y_3^-]gF^{\alpha n}(y_3^-)[y_3^-,y_1^-],
\label{Wilson_identity}
\eeq
where we used the step function
\beq
\theta(y_3^--y_2^-)=\int{dx\over 2\pi i}{e^{i(y_3^--y_2^-)x}\over x-i\epsilon}.
\eeq
We calculate each term in r.h.s of (\ref{Wilson_identity}) below.

\

\noindent
(1) first term

\beq
&&(p^+)^2\int{dy_1^-\over 2\pi}\int{dy_2^-\over 2\pi}e^{ix_1p^+y_1^-}e^{i(x_2-x_1)p^+y_2^-}
\la pS_{\perp}|\bar{\psi}_j(0)[0,y_1^-]D_{\perp}^{\alpha}(y_1^-)\psi_i(y_1^-)|pS_{\perp}\ra
\nonumber\\
&=&\delta(x_2-x_1)\Bigl[p^+\int{dy_1^-\over 2\pi}e^{ix_1p^+y_1^-}
\la pS_{\perp}|\bar{\psi}_j(0)[0,y_1^-]D_{\perp}^{\alpha}(y_1^-)\psi_i(y_1^-)|pS_{\perp}\ra\Bigr].
\label{first}
\eeq

\noindent
(2) second term

\beq
&&(p^+)^2\int{dy_1^-\over 2\pi}\int{dy_2^-\over 2\pi}\int^{\infty}_{y_1^-}dy_3^-
\,e^{ix_1p^+y_1^-}e^{i(x_2-x_1)p^+y_2^-}
\la pS_{\perp}|\bar{\psi}_j(0)[0,y_3^-]igF^{\alpha n}(y_3^-)[y_3^-,y_1^-]\psi_i(y_1^-)|pS_{\perp}\ra
\nonumber\\
&=&\delta(x_2-x_1)\Bigl[p^+\int{dy_1^-\over 2\pi}\int^{\infty}_{y_1^-}dy_3^-
\,e^{ix_1p^+y_1^-}\la pS_{\perp}|\bar{\psi}_j(0)[0,y_3^-]igF^{\alpha n}(y_3^-)[y_3^-,y_1^-]\psi_i(y_1^-)|pS_{\perp}\ra\Bigr].
\label{second}
\eeq

\noindent
(3) third term

\beq
&&-(p^+)^2\int{dy_1^-\over 2\pi}\int{dy_2^-\over 2\pi}\int dy_3^-\,\theta(y_3^--y_2^-)
e^{ix_1p^+y_1^-}e^{i(x_2-x_1)p^+y_2^-}
\la pS_{\perp}|\bar{\psi}_j(0)[0,y_3^-]igF^{\alpha n}(y_3^-)[y_3^-,y_1^-]\psi_i(y_1^-)|pS_{\perp}\ra
\nonumber\\
&=&-\int{dy_1^-\over 2\pi}\int{dy_2^-\over 2\pi}\int {dy_3^-\over 2\pi}\int dx
{(p^+)^2\over x-i\epsilon}e^{ix_1p^+y_1^-}e^{iy_3^-x}e^{i(\{x_2-x_1\}p^+-x)y_2^-}
\la pS_{\perp}|\bar{\psi}_j(0)[0,y_3^-]gF^{\alpha n}(y_3^-)[y_3^-,y_1^-]\psi_i(y_1^-)|pS_{\perp}\ra
\nonumber\\
&=&{1\over x_1-x_2+i\epsilon}\Bigl[p^+\int{dy_1^-\over 2\pi}\int {dy_3^-\over 2\pi}e^{ix_1p^+y_1^-}
e^{i(x_2-x_1)p^+y_3^-}
\la pS_{\perp}|\bar{\psi}_j(0)[0,y_3^-]gF^{\alpha n}(y_3^-)[y_3^-,y_1^-]\psi_i(y_1^-)|pS_{\perp}\ra\Bigr].
\label{third}
\eeq
Combining (\ref{first},\ref{second},\ref{third}), we can show 
$M^{\alpha}_D(x_1,x_2)={1\over x_1-x_2+i\epsilon}M^{\alpha}_F(x_1,x_2)+\delta(x_2-x_1)M^{\alpha}_{\partial}(x_1)$ and then the relation
among the twist-3 functions is given by
\beq
T_{q,D}(x_1,x_2)={1\over x_1-x_2+i\epsilon}T_{q,F}(x_1,x_2)+i\delta(x_2-x_1)f_{1T}^{\perp(1)}(x_1).
\eeq
Using the interchange symmetry $x_1\leftrightarrow x_2$, we can rewrite the above relation as
\beq
0&=&\Bigl({1\over x_1-x_2+i\epsilon}-{1\over x_1-x_2-i\epsilon}\Bigr)T_{q,F}(x_1,x_2)
+2i\delta(x_2-x_1)f_{1T}^{\perp(1)}(x_1).
\label{FDrelation}
\eeq
From the operator definition (\ref{Ftype}),
one can find that $T_{q,F}(x_1,x_2)$ contains the factor $e^{ix_1p^+(y_1^--y_2^-)}$. We can perform $x_1$-integration,
\beq
\int dx_1\Bigl({1\over x_1-x_2+i\epsilon}-{1\over x_1-x_2-i\epsilon}\Bigr)e^{ix_1p^+(y_1^--y_2^-)}
&=&-2\pi i\Bigl(\theta(y_2^--y_1^-)+\theta(y_1^--y_2^-)\Bigr)e^{ix_2p^+(y_1^--y_2^-)}
\nonumber\\
&=&-2\pi ie^{ix_2p^+(y_1^--y_2^-)},
\eeq
and then
\beq
\int dx_1\Bigl({1\over x_1-x_2+i\epsilon}-{1\over x_1-x_2-i\epsilon}\Bigr)T_{q,F}(x_1,x_2)
&=&-2\pi i T_{q,F}(x_2,x_2).
\eeq
After the integration of (\ref{FDrelation}) with respect to $x_1$, we can derive the relation
\beq
f_{1T}^{\perp(1)}(x)=\pi T_{q,F}(x,x),
\label{GIR2}
\eeq
which is nothing but the relation (\ref{GIR}).
This is well known relation between the first moment of the Sivers function $f_{1T}^{\perp(1)}(x)$
and the Qiu-Sterman function $T_{q,F}(x,x)$~\cite{Boer:2003cm,Ma:2003ut}. 
The same relation can be derived as we performed here in a simple way.
One can easily show the relation (\ref{LIR}) by the derivative of (\ref{GIR2}) with respect to $x$.

\section{Calculation of the derivative term ${\partial \over \partial k^{\alpha}}H(k)\Bigr|_{k=xp}$}

We show how to calculate the hard part ${\partial \over \partial k^{\alpha}}H(k)\Bigr|_{k=xp}$ in Eq.~(\ref{nonpolew_2}) without direct
operation of the $k$-derivative. We can calculate the part of the kinematical function as
\beq
i\int dx
\,{\rm Tr}[M^{\alpha}_{\partial}(x){\partial\over \partial k^{\alpha}}H(k)\Bigr|_{k=xp}]
&=&{M_N\over 2}\int {dx\over x}\epsilon^{\alpha \bar{n}nS_{\perp}}f^{\perp(1)}_{1T}(x)
{\rm Tr}[x\slash{p}{\partial\over \partial k^{\alpha}}H(k)\Bigr|_{k=xp}]
\nonumber\\
&=&{M_N\over 2}\int {dx\over x}\epsilon^{\alpha \bar{n}nS_{\perp}}f^{\perp(1)}_{1T}(x)\Bigl\{
{\partial\over \partial k^{\alpha}}{\rm Tr}[\slash{k}H(k)]\Bigr|_{k=xp}
-{\rm Tr}[\gamma^{\alpha}H(xp)]\Bigr\}.\hspace{4mm}
\label{k-part}
\eeq
We focus on the first term in the parenthesis. Because $H(k)$ carries the information about $k$, $q$ and $p_c$, it can be written by all possible Lorentz invariant variables,
\beq
{\rm Tr}[\slash{k}H(k)]=\hat{\sigma}(k^2,\tilde{s},\tilde{t},\hat{u},Q^2)
(2\pi)\delta(\tilde{s}+\tilde{t}+\hat{u}+Q^2-k^2),
\eeq
where we defined the variables, 
\beq
\tilde{s}=(k+q)^2,\hspace{5mm}\tilde{t}=(k-p_c)^2.
\eeq
We can set $k^2=0$ because 
${\partial\over \partial k^{\alpha}}k^2\Bigr|_{k=xp}{\partial\over\partial k^2}=2xp^{\alpha}{\partial\over\partial k^2}$ is canceled
with $\epsilon^{\alpha \bar{n}nS_{\perp}}$.
We find that $\hat{\sigma}(k^2,\tilde{s},\tilde{t},\hat{u},Q^2)$
coincides with $\hat{\sigma}(\hat s, \hat t, \hat u)$ in Eq. (\ref{twist-2}) in the collinear limit $k=xp$.
Then the $k$-derivative is converted into $\hat{s}$- and $\hat{t}$- derivatives,
\beq
{\partial\over \partial k^{\alpha}}[\hat{\sigma}(k^2,\tilde{s},\tilde{t},\hat{u},Q^2)
\delta(\tilde{s}+\tilde{t}+\hat{u}+Q^2-k^2)]
\Bigr|_{\tilde{s}=\hat{s},\tilde{t}=\hat{t}}
=\Bigl(2q^{\alpha}{\partial\over \partial\hat{s}}-2p_c^{\alpha}{\partial\over \partial\hat{t}}\Bigr)
[\hat{\sigma}(\hat s, \hat t, \hat u)\delta(\hat{s}+\hat{t}+\hat{u}+Q^2)].
\eeq
We calculate the $k$-derivative term in Eq. (\ref{k-part}) as
\beq
&&{M_N\over 2}\int {dx\over x}\epsilon^{\alpha \bar{n}nS_{\perp}}f^{\perp(1)}_{1T}(x){\partial\over \partial k^{\alpha}}{\rm Tr}[\slash{k}H(k)]\Bigr|_{k=xp}
\nonumber\\
&=&{\pi M_N}\int {dx\over x}\epsilon^{\alpha \bar{n}nS_{\perp}}f^{\perp(1)}_{1T}(x)
\Bigl(2q^{\alpha}{\partial\over \partial\hat{s}}-2p_c^{\alpha}{\partial\over \partial\hat{t}}\Bigr)
\hat{\sigma}(\hat s, \hat t, \hat u)\delta(\hat{s}+\hat{t}+\hat{u}+Q^2)
\nonumber\\
&=&{\pi M_N}\int {dx\over x}\epsilon^{\alpha \bar{n}nS_{\perp}}f^{\perp(1)}_{1T}(x)\Bigl\{
\delta(\hat{s}+\hat{t}+\hat{u}+Q^2)\Bigl(2q^{\alpha}{\partial\over \partial\hat{s}}-2p_c^{\alpha}{\partial\over \partial\hat{t}}\Bigr)
\hat{\sigma}(\hat s, \hat t, \hat u)
\nonumber\\
&&\hspace{40mm}+\Bigl({2q^{\alpha}-2p_c^{\alpha}\over 2p\cdot q-2p\cdot p_c}\Bigr)
\hat{\sigma}(\hat s, \hat t, \hat u){\partial\over \partial x}\delta(\hat{s}+\hat{t}+\hat{u}+Q^2)
\Bigr\}
\nonumber\\
&=&{\pi M_N}\int {dx\over x}\delta(\hat{s}+\hat{t}+\hat{u}+Q^2)\Bigl\{x{d\over dx}f^{\perp(1)}_{1T}(x)
\Bigl(\epsilon^{q\bar{n}nS_{\perp}}-\epsilon^{p_c\bar{n}nS_{\perp}}\Bigr){2\over \hat{u}}
\hat{\sigma}(\hat s, \hat t, \hat u)
\nonumber\\
&&+f^{\perp(1)}_{1T}(x)\Bigl[\Bigl(2\epsilon^{q\bar{n}nS_{\perp}}{\partial\over \partial\hat{s}}-2\epsilon^{p_c\bar{n}nS_{\perp}}
{\partial\over \partial\hat{t}}\Bigr)\hat{\sigma}(\hat s, \hat t, \hat u)
+\Bigl(\epsilon^{q\bar{n}nS_{\perp}}-\epsilon^{p_c\bar{n}nS_{\perp}}\Bigr){2\over \hat{u}}\Bigl(
x{\partial\over \partial x}\hat{\sigma}(\hat s, \hat t, \hat u)-\hat{\sigma}(\hat s, \hat t, \hat u)\Bigr)\Bigr]
\Bigr\}.
\eeq
We can calculate $x$-derivative of $\hat{\sigma}(\hat s, \hat t, \hat u)$ as
\beq
x{\partial\over \partial x}\hat{\sigma}(\hat s, \hat t, \hat u)=
\Bigl((\hat{s}+Q^2){\partial\over \partial \hat{s}}+\hat{t}{\partial\over \partial \hat{t}}\Bigr)
\hat{\sigma}(\hat s, \hat t, \hat u).
\eeq
Finally we combine the second term in Eq. (\ref{k-part}) and obtain the result in Eq. (\ref{nonpole}),
\beq
&&{\pi M_N}\int {dx\over x}\delta(\hat{s}+\hat{t}+\hat{u}+Q^2)\Bigl\{x{d\over dx}f^{\perp(1)}_{1T}(x)
\Bigl(\epsilon^{q\bar{n}nS_{\perp}}-\epsilon^{p_c\bar{n}nS_{\perp}}\Bigr){2\over \hat{u}}\hat{\sigma}(\hat s, \hat t, \hat u)
\nonumber\\
&&+f^{\perp(1)}_{1T}(x)\Bigl[\Bigl((\hat{s}+Q^2)\epsilon^{p_c\bar{n}nS_{\perp}}+\hat{t}\epsilon^{q\bar{n}nS_{\perp}}\Bigr){2\over \hat{u}}
\Bigl({\partial\over \partial\hat{t}}-{\partial\over \partial\hat{s}}\Bigr)\hat{\sigma}(\hat s, \hat t, \hat u)
-\Bigl(\epsilon^{q\bar{n}nS_{\perp}}-\epsilon^{p_c\bar{n}nS_{\perp}}\Bigr){2\over \hat{u}}\hat{\sigma}(\hat s, \hat t, \hat u)
\nonumber\\
&&-\epsilon^{\alpha \bar{n}nS_{\perp}}{\rm Tr}[\gamma_{\alpha}\bar{H}(xp)]
\Bigr]\Bigr\}.
\eeq
The derivative over the Mandelstam variable 
can be carried out after the calculation of the diagrams,
which is much easier than the direct $k$-derivative of $H(k)$.

%--------------------- Acknowledgment ----------

\section*{Acknowledgments}

The authors would like to thank Ivan Vitev and Matthew D. Sievert for fruitful discussion.
This research is supported by NSFC of China under Project No. 11435004 and research startup funding at SCNU.

\end{document}